\documentclass{aa}  
\usepackage{url}
\usepackage{graphicx}
\usepackage{txfonts}
\usepackage{amsmath}
\usepackage{subcaption}        
\usepackage{lscape}            
\usepackage{placeins}     

\begin{document}

   \title{A deep learning approach to member identification and mass estimation in galaxy clusters}

   \author{Zicheng Wang\inst{1}\email{lolola.ono@gmail.com}
        \and
        Liping Fu\inst{1}\corrauth{fuliping@shnu.edu.cn}
        \and
        Wei Du\inst{1}\corrauth{duwei@shnu.edu.cn}
        \and
        Wentao Luo\inst{2,3}\email{wtluo@ustc.edu.cn}
        \and
        Wenxiang Pei\inst{1}\email{wxpei@shnu.edu.cn}
        \and
        Zhu Chen\inst{1}\email{zhuchen@shnu.edu.cn}
        \and
        Zhijian Luo\inst{1}\email{zjluo@shnu.edu.cn}
        \and
        Yicheng Li\inst{4,1}\email{liyicheng@shao.ac.cn}
        \and
        Jiachen Qiu\inst{1}\email{jiachenq2141@163.com}
        \and
        Haoran Zhao\inst{1}\email{zhao-haoran@outlook.com}
        \and
        Kaiwen Mai\inst{1}\email{kaiwen_mai@163.com}
        \and
        Weiming Ren\inst{1}\email{1942045801@qq.com}
        }

    \institute{
    Shanghai Key Lab for Astrophysics, Shanghai Normal University, Shanghai 200234, China
    \and
    School of Aerospace Information, Hefei Institute of Technology, Zhizhong Road, Hefei, Anhui 238706, China
    \and
    Department of Astronomy, School of Physical Sciences, University of Science and Technology of China, Hefei, Anhui 230026, China
    \and
    Shanghai Astronomical Observatory, Chinese Academy of Sciences, Shanghai 200030, China
    }

   \date{}

  \abstract
   {
   Accurate dynamical mass estimates of galaxy clusters are essential for studies of cluster formation and evolution and for cosmological applications.
   However, interloper galaxies, which are not gravitationally bound to the cluster but are incorrectly identified as members, can bias dynamical mass measurements and increase their uncertainties.
   }
   {
   We developed a probabilistic deep-learning framework for cluster member identification and dynamical mass estimation and tested the sensitivity of the inferred masses to cluster miscentring and radially dependent incompleteness in controlled mock experiments.
   } 
   {
   We used mock galaxy cluster catalogues to train a two-stage model.
   In this model, a long short-term memory (LSTM) network assigns membership probabilities to galaxies arranged by projected phase-space proximity.
   These probabilities are then used to construct membership-weighted projected phase-space density maps, from which a Bayesian neural network (BNN) infers cluster masses and predictive uncertainties.
   }
   {
   The LSTM membership probabilities are well calibrated and form an isoprobability envelope in the projected phase space, with the $\langle p_{\rm mem}\rangle=0.5$ contour broadly tracing the empirical member--interloper transition.
   Applying a threshold of $p_{\rm th}=0.5$ provides a good balance between member completeness and purity, with high completeness within $R_{\rm proj}<r_{200}$.
   Membership weighting lowers the residual scatter of the fiducial BNN mass estimator, $\mathrm{BNN}_{\mathrm{2D}}^{\mathrm{Mem}}$, to $\sigma_\epsilon=0.120$, a $19.5\%$ reduction relative to the unweighted baseline, although the predictive uncertainties are not uniformly calibrated across the BNN models.
   Using the brightest cluster galaxy rather than the gravitational centre lowers the median inferred mass by up to 47.6\%, with larger reductions at larger centring offsets. By contrast, removing 40\% of eligible non-central galaxies within $r_{200}$ lowers it by 6.1\% relative to complete catalogues.
   }
   {}


   \maketitle
   \nolinenumbers

\section{Introduction}
As the most massive gravitationally bound structures in the Universe, galaxy clusters occupy the nodes of the cosmic web. Their gravitational potentials are dominated by dark matter, while galaxies and the intracluster medium provide observable tracers of the underlying mass distribution. Accurate measurements of cluster masses and mass distributions are therefore essential for studying cluster formation and evolution and for constraining cosmological parameters \citep{Schuecker2002The, Rozo_2010, Wen2010Mass, Falco2013A, planck2014}. Owing to their broad astrophysical and cosmological significance, galaxy clusters are major science targets of successive generations of wide-field surveys, including the Sloan Digital Sky Survey (SDSS) \citep{SDSSDR18,SDSSClusterWen}, the Dark Energy Survey (DES) \citep{DESDR2,DESCluster}, the Euclid survey \citep{EuclidOverview,EuclidCluster}, the Vera C. Rubin Observatory's Legacy Survey of Space and Time (LSST) \citep{LSST}, and the upcoming China Space Station Telescope (CSST) survey \citep{CSST2025,CSSTCluster}.

Various observational probes have been developed to constrain galaxy cluster masses. Gravitational lensing probes the projected mass distribution through the distortion of background galaxy images by the cluster gravitational field \citep{Du_2015,Du_2014,Hoekstra2013Masses}. X-ray observations of the hot intracluster medium can be used to infer cluster mass profiles, commonly under the assumption of hydrostatic equilibrium \citep{XrayMassprofile}. The Sunyaev--Zel'dovich effect, arising from inverse Compton scattering between cosmic microwave background photons and hot electrons in the intracluster medium, provides another important mass proxy \citep{SZmassrelation}. Alternatively, the positions and line-of-sight (LOS) velocities of cluster galaxies can be used to trace the cluster gravitational potential through their dynamics. Such dynamical approaches are particularly valuable for probing cluster mass distributions to large cluster-centric radii, where some gas-based observables become increasingly difficult to measure \citep{DynamicsofGCs}.

To estimate cluster masses from member galaxies, a straightforward approach is to apply scaling relations between the total mass enclosed within a characteristic radius and the velocity dispersion of member galaxies, which are physically motivated by the virial theorem \citep{VirialMunari,virial,MassSigma}. Under assumptions about the tracer density and velocity anisotropy profiles, the cluster mass distribution can also be constrained through the Jeans equation \citep{Jeanseq}. Unlike methods based primarily on galaxy velocity dispersions, the caustic technique \citep{Diaferio2005Caustic} exploits the projected phase-space distribution of cluster galaxies, defined by projected cluster-centric radius and LOS velocity, to estimate the radial escape-velocity profile and thereby infer the cluster mass profile \citep{Serra2011}. Although these approaches provide valuable constraints on cluster masses, their accuracy can be affected by assumptions about the dynamical state, velocity anisotropy, geometry, or mass distribution of clusters. A common challenge for galaxy-based dynamical methods is interloper contamination, in which foreground or background galaxies projected along the LOS are mistakenly identified as cluster members.

Interlopers can distort the inferred dynamical properties and projected phase-space distribution of galaxy clusters, thereby biasing dynamical mass estimates \citep{Biviano,Wojtak2007,Ferragamo}. Additional galaxy information and more sophisticated selection methods can improve the separation of cluster members from interlopers. For example, galaxy positions combined with single-band photometric information can be used to identify galaxy clusters and estimate their richness \citep{voronoi}. With multi-band photometric data, galaxy membership probabilities can be estimated more accurately, enabling the construction of cluster catalogues with improved purity and completeness \citep{redMaPPer4}. When spectroscopic information is available, projected phase-space information provides additional leverage for identifying cluster members. For example, \citet{Serra2013} showed that the caustic method, which uses galaxy positions and LOS velocities, can effectively separate interlopers from cluster members.

Recently, machine-learning and deep-learning techniques have emerged as data-driven alternatives for dynamical cluster mass estimation, allowing cluster masses to be inferred directly from galaxy phase-space information. For instance, \citet{Ntampaka} introduced a support distribution machine (SDM) that learns from the joint distribution of galaxy LOS velocities and projected cluster-centric distances, reducing the 68\% scatter of the fractional mass error from $0.87$ for the standard $M$--$\sigma_v$ scaling relation to $0.46$. \citet{CNN2D} subsequently developed convolutional neural network (CNN) estimators using LOS velocities alone or together with projected radii, achieving mass-estimation accuracies of $0.174$ dex and $0.132$ dex, respectively. Approximate Bayesian extensions of these models were later developed to characterise predictive uncertainties \citep{Uncertainties}, while \citet{Kodi2021} further incorporated 2D projected galaxy positions and LOS velocities using a 3D CNN. These studies demonstrate the effectiveness of machine learning for dynamical cluster mass estimation. However, interloper contamination remains a challenge in these data-driven approaches. Given the ability of machine-learning methods to extract complex patterns from galaxy phase-space distributions, they may also help address this issue.

In this study, we developed a deep-learning framework that incorporates probabilistic galaxy membership into dynamical cluster mass inference, with the aim of reducing the impact of interloper contamination and improving mass-estimation precision. Following the methodology of \citet{CNN2D}, we generated mock observational data for galaxy clusters. Models based on long short-term memory (LSTM) networks have been applied to a variety of astronomical problems, including broad absorption line detection in quasar spectra \citep{LSTM-LYY} and photometric redshift estimation from wavelength-ordered, multi-band flux sequences \citep{LSTM-EGG,LSTM-Chen}. In our study, we constructed an LSTM network \citep{Hochreiter1997} to estimate individual galaxy membership probabilities, with galaxies ordered according to their proximity in projected phase space. The predicted membership probabilities were subsequently incorporated into the projected phase-space distribution used by a Bayesian neural network (BNN) to infer cluster masses and their predictive uncertainties. We further investigated the role of galaxy magnitude information in both membership identification and cluster mass estimation.

To approximate realistic observational conditions, we imposed an $r$-band AB apparent-magnitude limit of 21 on the mock galaxy catalogues. We further assessed the robustness of our framework through two additional tests, examining the effects of cluster miscentring and radially dependent data incompleteness on mass inference. Finally, as an initial application to observed systems, we applied our method to a sample of observed clusters with SDSS spectroscopic data and compared the inferred masses with literature estimates from weak-lensing, X-ray, and dynamical analyses.

The remainder of this paper is structured as follows. In Sect.~\ref{sec:Data}, we describe the generation of the datasets used for training and testing. Section~\ref{sec:Method} introduces the LSTM and BNN models and describes their architectures and training procedures. In Sect.~\ref{sec:result}, we present our results. We summarise our main conclusions in Sect.~\ref{sec:conclusion}.

\section{Data}
\label{sec:Data}
    
\subsection{Mock galaxy catalogue}
\label{sec:cosmo} 
We extracted mock galaxies from the catalogue of \citet{Pei2024}, who applied a modified version of the semi-analytic model \textsc{L-Galaxies} \citep{LGalaxy} to JiuTian-1G (hereafter JT1G), a large-volume dark-matter-only $N$-body simulation. The JT1G simulation contains $6144^3$ particles within a cubic box with a side length of $1\,h^{-1}\mathrm{Gpc}$. The simulation adopts the cosmological parameters $\Omega_\Lambda = 0.6889$, $\Omega_m = 0.3111$, $\Omega_b = 0.0490$, $h = 0.6766$ and $\sigma_8 = 0.8102$ \citep{Planck2020}. It traces particles from redshift $z = 127$ to $z = 0$, and stores $128$ snapshots with an average time interval of approximately 100~Myr between adjacent snapshots. The mass resolution of this simulation is $3.72\times10^8 h^{-1}M_\odot$. Dark matter haloes and substructures in JT1G are identified using the friends-of-friends (FoF) and \textsc{SUBFIND} algorithms \citep{Fof,GADGET2}.

\citet{Pei2024} improved the \textsc{L-Galaxies} model \citep{LGalaxy} by addressing its time convergence problem. By applying the modified \textsc{L-Galaxies} model to the JT1G simulation, they generated mock galaxy catalogues that successfully reproduce the stellar mass function out to redshift $z \simeq 3$ and meet the requirements for realistic forecasts for next-generation large-scale surveys such as Euclid, LSST, and CSST. In this work, we adopted the galaxy catalogue with CSST $r$-band magnitudes and selected relatively bright galaxies to achieve a sample depth comparable to that expected for future spectroscopic galaxy surveys.

\subsection{Mock cluster observations}
\label{sec:Mock}
We followed the methodology introduced in \citet{CNN2D} to generate mock cluster catalogues using the simulation data described in Sect.~\ref{sec:cosmo}. Specifically, we adopted the simulation data from snapshot 116, which corresponds to a redshift of $z\simeq0.13$. We selected FoF haloes with mass $M_{200} \geq 10^{13.5} h^{-1} M_{\odot}$ and subhalo richness $N \geq 10$ as candidate clusters, where $M_{200}$ denotes the mass enclosed by a radius $r_{200}$ within which the mean density is 200 times the critical density of the Universe. We adopted the gravitational centre, defined as the position of minimum gravitational potential, as the `true' cluster centre.
Galaxies were selected as subhaloes with stellar mass $M_{\star} \geq 10^{9.5}\,h^{-1}M_{\odot}$, and the subhalo richness $N$ was computed using galaxies satisfying this stellar-mass cut. This selection yields 80,849 cluster candidates at $z\simeq0.13$, from which the training and test samples are drawn. To avoid missing galaxies for clusters close to the boundary of the periodic simulation box, we constructed a periodically padded galaxy catalogue by copying slices of galaxies from the opposite sides of the box. The padding width is $120~h^{-1}{\rm Mpc}$ on each face.

For each mock cluster observation, we assigned a random LOS and placed a mock observer such that the cluster lies at the comoving distance corresponding to the snapshot redshift $z\simeq0.13$. The LOS unit vector $\hat{z}_{\rm los}$ is defined from the observer to the cluster centre, and two orthonormal basis vectors perpendicular to it are denoted by $\hat{x}_{\rm los}$ and $\hat{y}_{\rm los}$. We calculated the projected positions of galaxies relative to the gravitational centre of the target cluster as
\begin{align}\label{eq:rproject}
  x_{\rm proj} &=
  \left[\boldsymbol{r}-(\boldsymbol{r}\cdot\hat{z}_{\rm los})\hat{z}_{\rm los}\right]\cdot\hat{x}_{\rm los}, \\
  y_{\rm proj} &=
  \left[\boldsymbol{r}-(\boldsymbol{r}\cdot\hat{z}_{\rm los})\hat{z}_{\rm los}\right]\cdot\hat{y}_{\rm los},
\end{align}
where $\boldsymbol{r}$ is the comoving position vector from the cluster gravitational centre to the target galaxy. The projected radius is then
$R_{\rm proj}=(x_{\rm proj}^2+y_{\rm proj}^2)^{1/2}$. We retained galaxies within a fixed projected aperture of $R_{\rm proj}\leq 4~h^{-1}{\rm Mpc}$. This aperture was chosen to be large enough to include the cluster outskirts and the nearby projected environment, where infalling galaxies and interlopers are expected to appear across different systems.

For galaxies satisfying the projected-radius cut, we computed the LOS velocity relative to the cluster centre as
\begin{equation}
    v_{\rm los}
    =
    \left[
    v_{\rm gal}^{(\rm P,los)}-v_{\rm cen}^{(\rm P,los)}
    \right]
    +
    \left[
    v_{\rm gal}^{(\rm H)}-v_{\rm cen}^{(\rm H)}
    \right],
    \label{eq:vlos}
\end{equation}
where $v^{(\rm P,los)}$ is the peculiar velocity projected along the LOS, and $v^{(\rm H)}$ is the Hubble-flow velocity inferred from the comoving distance to the observer. The subscripts `gal' and `cen' denote the target galaxy and the cluster centre, respectively. We further required $|v_{\rm los}|\leq 3000~{\rm km}~{\rm s}^{-1}$. This relatively broad velocity window was chosen to retain high-velocity cluster members and galaxies in the infall region, while excluding very distant foreground and background galaxies. Together with the projected-radius cut, this selection defines a broad projected phase-space window containing both members and interlopers, allowing the network to learn membership classification in the presence of projection-induced contamination. Galaxies were labelled as members if they belong to the same FoF group as the target cluster and as interlopers otherwise. The FoF membership of galaxies is based on spatial linking and does not guarantee that all members are gravitationally bound to the cluster \citep{Fof,Bett2007}.

\begin{figure*}[t]
    \centering
    \includegraphics[width=\textwidth]{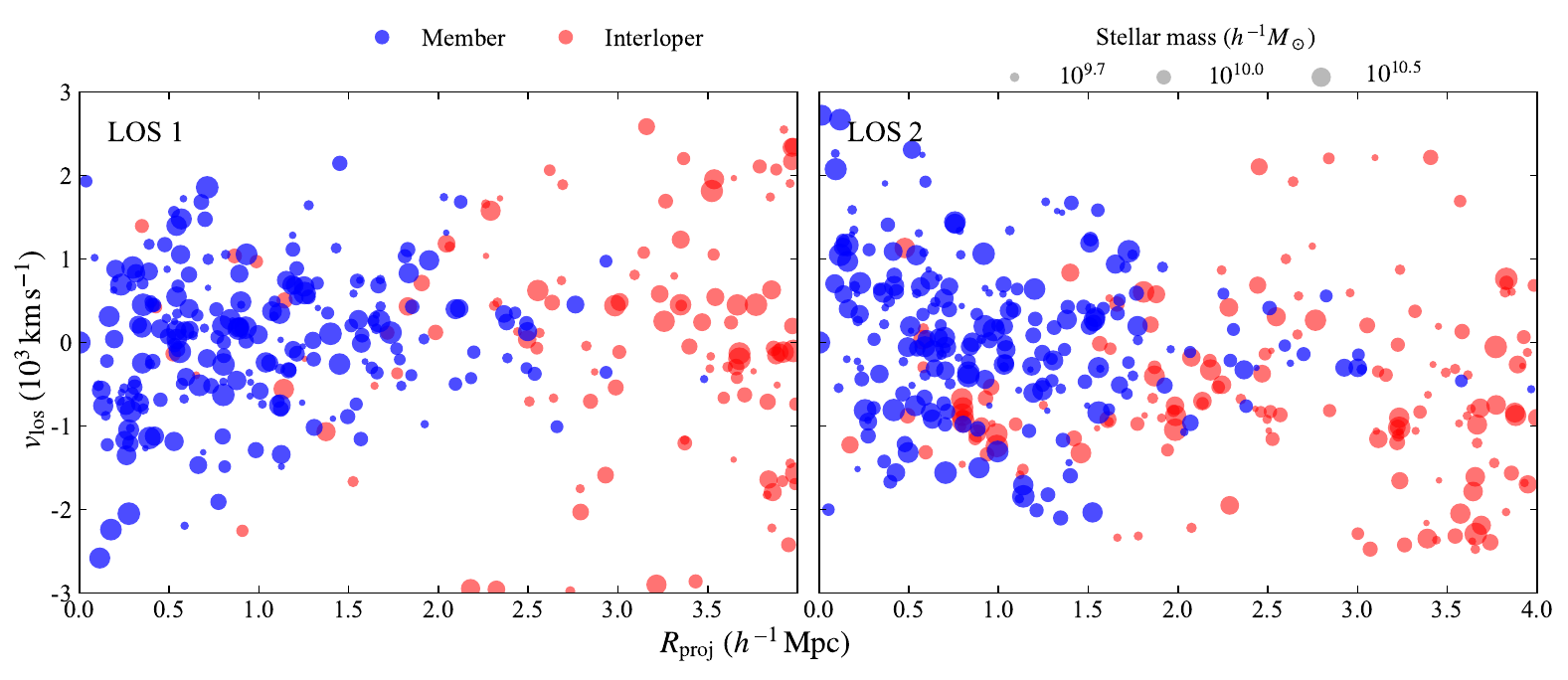}
    \caption{Projected phase-space distributions of galaxies in a single galaxy cluster viewed along two different lines of sight. The blue and red points denote cluster members and interlopers, respectively. The marker size encodes the stellar mass, $M_\star$, with the larger symbols corresponding to more massive galaxies.}
    \label{fig:mocksample}
\end{figure*}

Figure~\ref{fig:mocksample} illustrates the projected phase-space distributions of galaxies in a single galaxy cluster viewed along two different lines of sight. The figure shows that the apparent distribution of galaxies in the $R_{\rm proj}$--$v_{\rm los}$ plane depends on the LOS projection. Interlopers can occupy a broad region of the projected phase space, thereby complicating the identification of cluster members and introducing additional uncertainties into cluster mass estimation. This motivates the use of auxiliary galaxy information to mitigate the impact of interlopers on mass inference. We therefore tested whether galaxy magnitude and inferred membership probability provide additional predictive information for cluster mass estimation.

\begin{figure*}

	\includegraphics[width=\linewidth]{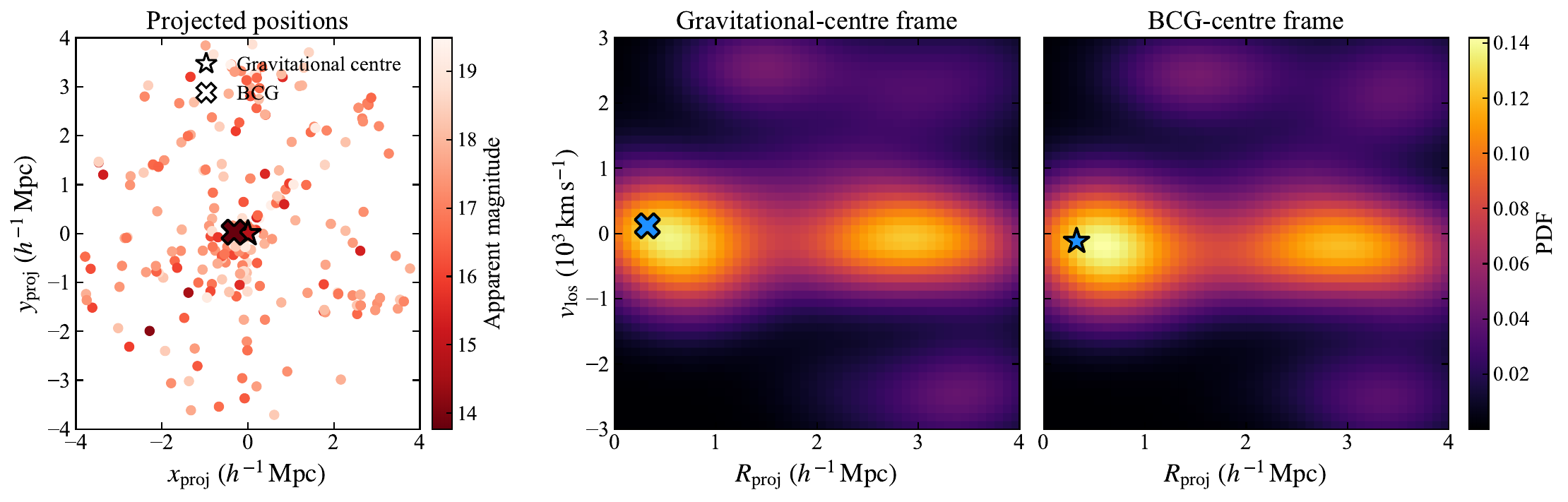}
    \caption{Effect of the adopted cluster centre on the projected phase-space distribution. Left: Projected positions of galaxies in the field of view, with the points coloured by their $r$-band apparent magnitude, capped at $m_r=21$. The star marks the gravitational centre, defined as the position of minimum gravitational potential within the cluster, while the cross marks the BCG. Middle and right: Gaussian KDEs of the projected phase-space PDF in the $R_{\rm proj}$--$v_{\rm los}$ plane, using the gravitational centre and the BCG as the adopted centre, respectively. Both cluster members and interlopers are included in the KDE estimates.}

    \label{fig:Obs_sample}
\end{figure*}

To mimic observational limitations, we constructed a dataset with an $r$-band AB apparent-magnitude limit of 21, corresponding to a brighter cut than the nominal SDSS $r$-band depth for point sources of $\sim 22.2$ \citep{SDSSrbandlimit}. The left panel of Fig.~\ref{fig:Obs_sample} illustrates the distribution of galaxies based on their projected positions, with each point representing a galaxy and coloured by its apparent magnitude.

\subsection{Dataset}
\label{sec:Dataset}

\begin{figure}

	\includegraphics[width=\columnwidth]{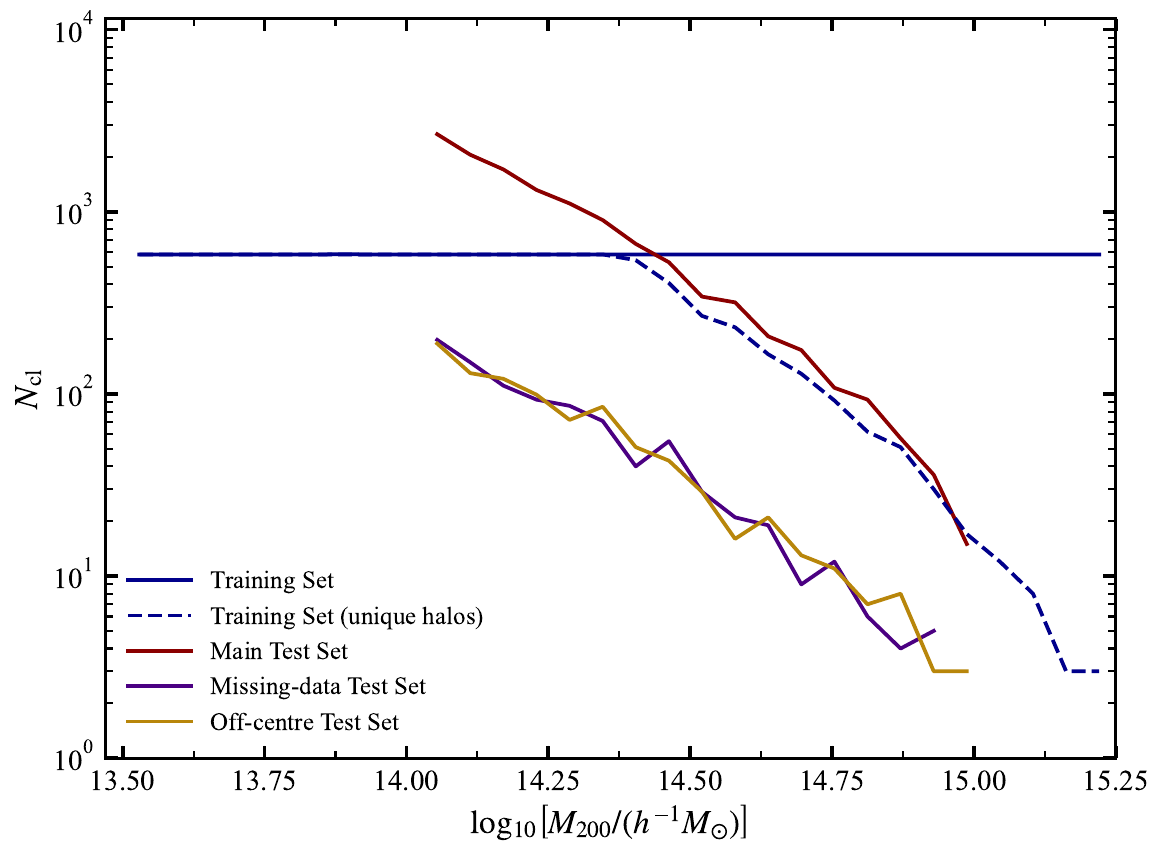}
    \caption{Mass distributions of the datasets. The solid curves show the number of mock cluster realisations in each mass bin, while the dashed dark-blue curve shows the number of unique haloes contributing to the training set. The same 30 equal-width bins spanning $13.5$--$15.25$ in $\log_{10}[M_{200}/(h^{-1}M_\odot)]$ are used for all datasets, with only non-empty bins shown. For the training set, an approximately flat mass distribution is constructed by varying both the number of selected haloes and the number of random LOS realisations per halo across the mass range. The missing-data and off-centre test sets are described in Sect.~\ref{sec:Dataset}.}
    
    \label{fig:MF}
\end{figure}

To reduce sensitivity to the underlying cluster mass function during training, we constructed a training set with an approximately flat mass distribution over the range $10^{13.5}$--$10^{15.25}\,h^{-1}M_\odot$, following \citet{CNN2D}. In the mass range $10^{14}$--$10^{15}\,h^{-1}M_\odot$, we randomly assigned $70\%$ of the cluster candidates to the training set and reserved the remaining $30\%$ for the main test set. This cluster-level split ensures that different LOS projections of the same underlying cluster do not appear in both the training and test samples \citep{NoF}. To obtain an approximately flat training set distribution in $\log_{10} M_{200}$, we divided the logarithmic mass range into 30 equal-width bins, targeting approximately 583 mock realisations per bin. We used one random LOS per selected halo in well-populated bins with $M_{200} < 10^{14.37}\,h^{-1}\,M_\odot$, and up to approximately 195 LOS realisations per halo in the highest-mass bin. 
Multiple LOS projections increase the diversity of viewing orientations and help prevent the training set from being dominated by the more abundant low-mass clusters. However, projections of the same halo are not independent and do not increase the halo-to-halo diversity at the high-mass end. The cluster-level split described above prevents information leakage arising from projections of the same halo appearing in both the training and main test sets. Nevertheless, the limited number of independent high-mass clusters may limit the generalisability of predictions for the rarest systems.
For the main test set, each cluster in the reserved $30\%$ sample was projected along the three orthogonal Cartesian axes ($x$, $y$, and $z$), yielding three mock realisations per cluster. For each cluster projection, galaxies were labelled as either members or interlopers according to their FoF group membership.

Accurately identifying the centre of a galaxy cluster is challenging because different centre definitions, such as the position of the brightest cluster galaxy (BCG), the X-ray peak, and the minimum of the gravitational potential, do not always coincide. Within the mass range of the main test set, 6,335 cluster candidates have an $r$-band BCG that does not coincide with the galaxy located at the gravitational centre. To quantify the impact of miscentring on model performance, we randomly selected 1,000 of these clusters and generated paired mock observations for each: one centred on the gravitational centre and the other adopting the $r$-band BCG as the cluster centre. These paired samples constitute the off-centre test set. Figure~\ref{fig:Obs_sample} illustrates how this recentring affects the projected phase-space distribution of galaxies in one example cluster. In this case, the BCG is offset from the gravitational centre by $0.324\,h^{-1}{\rm Mpc}$ in projection. The two right-hand panels show that the adopted centre may change the structure in the phase-space plane, potentially biasing the inferred cluster mass.

In addition to centre offsets, incomplete galaxy sampling caused by blending, instrumental limitations, observational strategies, or other observational constraints can also affect the accuracy of the network predictions. To quantify the impact of data incompleteness, we randomly selected 1,000 clusters from the main test set and constructed missing-data test sets by removing fixed fractions of galaxies from specified radial annuli. Specifically, we considered missing fractions of 10\%, 20\%, and 40\%,
applied separately within the projected-radius intervals
$0 < R_{\rm proj}/r_{200} < 1$,
$1 < R_{\rm proj}/r_{200} < 2$, and
$2 < R_{\rm proj}/r_{200} < 3$. For each cluster, this yields nine missing-data realisations, one for each combination of missing fraction and projected radial interval. The mass distributions of the different datasets are shown in Fig.~\ref{fig:MF}.

In summary, we constructed a training set containing 17,489 mock realisations, a main test set containing 13,569 mock realisations generated from 4,523 distinct underlying clusters, an off-centre test set containing 1,000 clusters to assess the impact of cluster miscentring, and a set of missing-data test samples to evaluate the effect of radial incompleteness under different missing fractions.

\section{Method}
\label{sec:Method}

In the Lambda cold dark matter ($\Lambda$CDM) paradigm, galaxy clusters assemble hierarchically and grow approximately inside-out \citep{GCsformation}. As a result, galaxies located closer to the cluster centre are, in a statistical sense, more likely to have been accreted earlier, while genuine member galaxies tend to occupy coherent structures in projected phase space under the cluster gravitational potential. Motivated by this physical picture, we represented each cluster as an ordered pseudo-sequence, starting from the central galaxy and proceeding according to phase-space proximity. Although this sequence is not a true temporal sequence, it encodes local continuity in projected position--velocity space, allowing the LSTM to exploit contextual information among neighbouring galaxies when assigning membership probabilities. The resulting probabilistic member sample was then used for downstream cluster mass inference.

\subsection{Long short-term memory (LSTM)}
\label{sec:lstm}

\begin{figure*}

	\includegraphics[width=\linewidth]{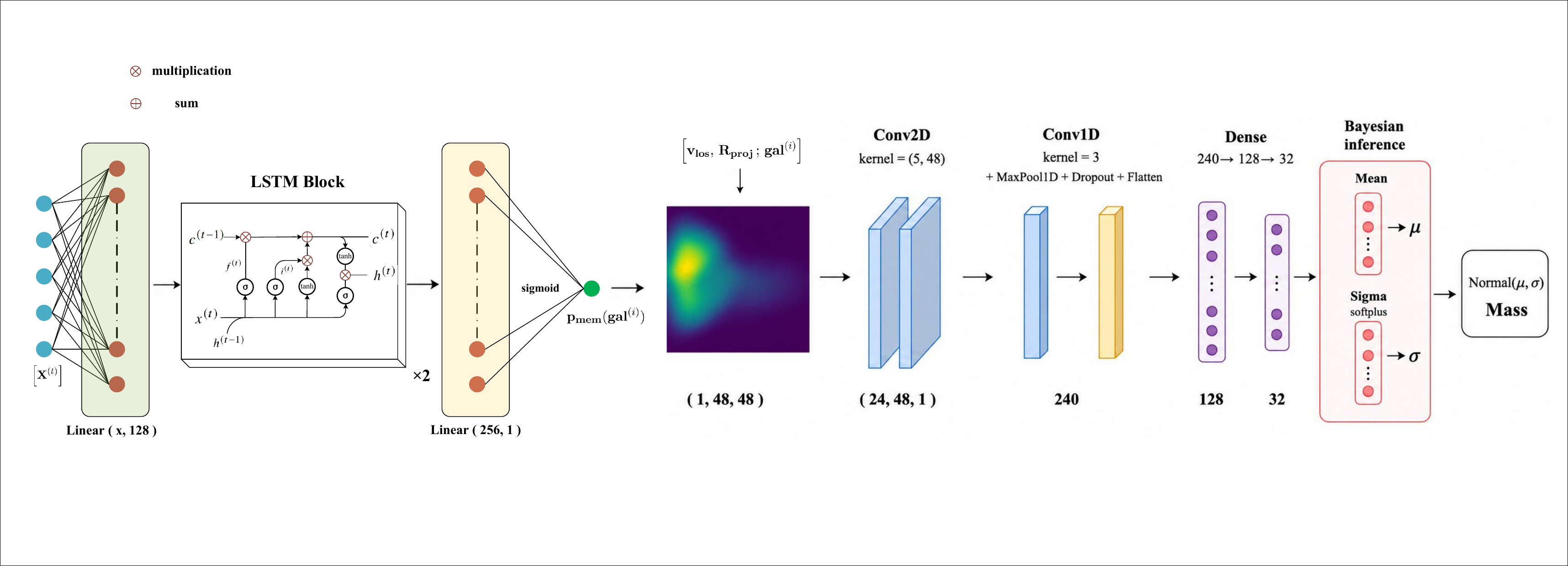}
    \caption{
    Schematic of the model architecture for membership-probability prediction and mass estimation.
    The BNN uses the membership-weighted phase-space PDF in $(v_{\rm los}, R_{\rm proj})$ as input, with Bayesian inference applied only at the final output stage through two Bayesian linear layers.
    }
    \label{fig:model_architecture}
\end{figure*}

Long short-term memory networks are recurrent neural networks designed to model sequential data by regulating information flow through gated memory cells \citep{Hochreiter1997, LSTMReview}. For an input sequence $\{x^{(t)}\}$, an LSTM updates the hidden state $h^{(t)}$ and cell state $c^{(t)}$ through input, forget, and output gates:

\begin{flalign}
& i^{(t)}=\sigma \left( W_{i,x} x^{(t)}+W_{i,h} h^{(t-1)}+b_i \right), 
\label{eq:InputGate}&\\[5pt]
& f^{(t)}=\sigma \left( W_{f,x} x^{(t)}+W_{f,h} h^{(t-1)}+b_f \right),
\label{eq:ForgetGate}&\\[5pt]
& \tilde{c}^{(t)}=\tanh \left( W_{c,x} x^{(t)}+W_{c,h} h^{(t-1)}+b_c \right),
\label{eq:CandidateCell}&\\[5pt]
& c^{(t)}=f^{(t)} \odot c^{(t-1)}+i^{(t)} \odot \tilde{c}^{(t)},
\label{eq:Cell}&\\[5pt]
& o^{(t)}=\sigma \left( W_{o,x} x^{(t)}+W_{o,h} h^{(t-1)}+b_o \right),
\label{eq:OutputGate}&\\[5pt]
& h^{(t)} = o^{(t)} \odot \tanh \left(c^{(t)}\right).
\label{eq:HiddenState}&
\end{flalign}

Here, $\sigma$ is the sigmoid activation function, $\odot$ denotes element-wise multiplication, and the weights $W$ and biases $b$ are learnt during training. The candidate cell state $\tilde{c}^{(t)}$ is an internal quantity of the standard LSTM update. 
 
In our application, $x^{(t)}$ represents the feature vector of the $t$-th galaxy in the ordered pseudo-sequence. The LSTM therefore assigns membership probabilities using not only the features of individual galaxies, but also contextual information from neighbouring galaxies in the sequence. Since genuine cluster members are expected to form coherent structures in projected phase space, whereas interlopers are less strongly constrained by the cluster potential, this sequential representation was designed to provide contextual information that may help the network distinguish members from contaminants.

The hidden-state updates of an LSTM depend on the order of the input sequence, making the galaxy ordering an important part of the model construction. To construct the input sequence, we adopted a simple iterative nearest-neighbour sorting algorithm in projected phase space. The sequence started from the central galaxy, and each subsequent galaxy was selected as the nearest remaining galaxy to the previously selected one.

We normalised the projected coordinates by the aperture radius, $R_{\rm cut}=4\,h^{-1}{\rm Mpc}$, as well as the LOS velocity using a robust median absolute deviation (MAD) estimate. Specifically, we define
\begin{equation}
\xi = \frac{x_{\rm proj}}{R_{\rm cut}},
\qquad
\eta = \frac{y_{\rm proj}}{R_{\rm cut}},
\qquad
u_{\rm los} =
\frac{v_{\rm los}-\tilde{v}_{\rm los}}{\sigma_{\rm los}},
\label{eq:scaled_features}
\end{equation}
where $\xi$ and $\eta$ are the normalised projected coordinates, $u_{\rm los}$ is the normalised LOS velocity, and $\tilde{v}_{\rm los}$ is the median LOS velocity of galaxies in the candidate cluster. The MAD-based LOS velocity scale is defined as
\begin{equation}
\sigma_{\rm los}
=
\max \left[
1.4826\,{\rm median}\left(\left|v_{\rm los}-\tilde{v}_{\rm los}\right|\right),
300\,{\rm km\,s^{-1}}
\right].
\label{eq:mad_velocity}
\end{equation}
The lower limit of $300\,{\rm km\,s^{-1}}$ was introduced to avoid unstable velocity normalisation when the MAD-based velocity scale is very small, especially in low-richness systems.

The phase-space distance between two galaxies was then computed as
\begin{equation}
d_{12}
=
\sqrt{
\left(\xi_2-\xi_1\right)^2
+
\left(\eta_2-\eta_1\right)^2
+
\left(u_{{\rm los},2}-u_{{\rm los},1}\right)^2
}.
\label{eq:Distance}
\end{equation}
At each step, the remaining galaxy with the smallest phase-space distance from the last selected galaxy was appended to the sequence. To assess the effect of this sequence-construction choice, we also trained an otherwise identical LSTM using randomly ordered galaxy sequences, while retaining the same galaxy samples and input features. We compare the resulting membership predictions and downstream mass-estimation performance with those obtained using the fiducial ordering in Sect.~\ref{sec:massestimation}.

For the galaxy at sequence position $t$, the LSTM maps an input feature vector 
$\mathbf{X}^{(t)}$ to a membership probability $p_{\rm mem}^{(t)}$. We grouped the input features as
\begin{equation}
\mathbf{X}^{(t)}
=
\left[
\mathbf{X}_{\rm pos},
\mathbf{X}_{\rm vel},
\mathbf{X}_{\rm cl},
\mathbf{X}_{\rm den},
\mathbf{X}_{\rm mag}
\right]^{(t)}.
\label{eq:lstm_input_groups}
\end{equation}

The position-related features, $\mathbf{X}_{\rm pos}$, describe the projected location of each galaxy within the cluster-centred aperture. They include the normalised projected coordinates, $\xi$ and $\eta$, the aperture-normalised projected radius $\rho=R_{\rm proj}/R_{\rm cut}$, and the relative projected radius $r_{50}=R_{\rm proj}/R_{50,\rm obs}$, where $R_{50,\rm obs}$ is the median projected radius of galaxies in the same aperture. The feature $r_{50}$ measures the radial position of each galaxy relative to the observed radial scale of its candidate cluster.

The velocity-related features, $\mathbf{X}_{\rm vel}$, describe both the LOS velocity coordinate of each galaxy within the fixed velocity window and its velocity relative to the candidate cluster. We define the fixed-aperture normalised LOS velocity as
\begin{equation}
\hat{v}_{\rm los}
=
\frac{v_{\rm los}}{v_{\rm cut}},
\qquad
v_{\rm cut}=3000\,{\rm km\,s^{-1}}.
\label{eq:fixed_velocity}
\end{equation}
The fixed-aperture velocity $\hat{v}_{\rm los}$ preserves the velocity-space position of each galaxy within the common LOS velocity window, whereas $u_{\rm los}$ measures its velocity offset relative to the median and internal velocity scale of the candidate cluster. We define the projected phase-space distance proxy as
\begin{equation}
d_{\rm ps}
=
\sqrt{\rho^2+\hat{v}_{\rm los}^{\,2}}.
\label{eq:dps_feature}
\end{equation}
We also included the dimensionless velocity-scale feature
$s_{\rm los}=\sigma_{\rm los}/v_{\rm cut}$. The velocity-related input group is therefore
\begin{equation}
\mathbf{X}_{\rm vel}
=
\left[
\hat{v}_{\rm los},
\left|\hat{v}_{\rm los}\right|,
d_{\rm ps},
u_{\rm los},
\left|u_{\rm los}\right|,
s_{\rm los}
\right].
\end{equation}

To provide global information on the candidate cluster, we used the cluster-level observable feature vector $\mathbf{X}_{\rm cl}
=
\left[
\log_{10} N_{\rm gal},
f_{<1},
f_{1-2}
\right]$,
where $N_{\rm gal}$ is the number of galaxies in the cluster-centred aperture, and 
$f_{<1}$ and $f_{1-2}$ are the fractions of galaxies with 
$R_{\rm proj}<1\,h^{-1}{\rm Mpc}$ and 
$1\,h^{-1}{\rm Mpc} \leq R_{\rm proj}<2\,h^{-1}{\rm Mpc}$, respectively. These quantities summarise the richness and projected radial concentration of the candidate cluster, and were repeated for all galaxies in the same aperture when concatenated with the per-galaxy features at each sequence position.

The local-density features describe the local environment of each galaxy within the cluster-centred aperture. Genuine cluster members are expected to be embedded in the main cluster structure in projected phase space, whereas interlopers may appear as isolated objects or as part of distinct foreground/background structures. Thus, local density quantifies how strongly a galaxy is connected to a coherent phase-space environment, beyond its individual projected position and LOS velocity. For each galaxy, we computed $k$-nearest-neighbour density estimates in two spaces, with 
$n^{\rm ps}_{k}$ measuring the number density in the normalised projected phase space $(\xi,\eta,u_{\rm los})$ and $\Sigma^{\rm xy}_{k}$ measuring the surface number density 
in the projected coordinate plane $(\xi,\eta)$. Here, $k$ sets the neighbour scale, with smaller values probing compact local associations, while larger values trace more extended cluster-scale environments. Let $d^{\rm ps}_{k}$ and $d^{\rm xy}_{k}$ be the distances to the $k$th nearest neighbour in these two spaces. We define
\begin{equation}
n^{\rm ps}_{k}
=
\frac{k}{(4\pi/3)(d^{\rm ps}_{k}+\epsilon)^3},
\qquad
\Sigma^{\rm xy}_{k}
=
\frac{k}{\pi(d^{\rm xy}_{k}+\epsilon)^2},
\end{equation}
where $\epsilon$ prevents numerical divergence when the nearest-neighbour distance is very small. The density input vector is $\mathbf{X}_{\rm den}
=
\left[
\log_{10} n^{\rm ps}_{5},
\log_{10} n^{\rm ps}_{10},
\log_{10} n^{\rm ps}_{20},
\log_{10} \Sigma^{\rm xy}_{10}
\right]$.
Together, the different choices of $k$ and density spaces allow $\mathbf{X}_{\rm den}$
to characterise the galaxy environment hierarchically, from compact local
associations to broader cluster-scale structures in both projected phase space and the sky plane.

The magnitude-related feature is the standardised rest-frame $r$-band absolute magnitude,
\begin{equation}
\mathbf{X}_{\rm mag}=[\widetilde{M}_r],
\qquad
\widetilde{M}_r=
\frac{\langle M_r\rangle-M_r}{\sigma_{M_r}},
\end{equation}
where $M_r$ is the rest-frame $r$-band AB absolute magnitude, and
$\langle M_r\rangle$ and $\sigma_{M_r}$ were computed using galaxies within
the same cluster-centred aperture. This feature measures how much brighter or fainter
a galaxy is relative to the mean magnitude scale of its aperture. We also trained a magnitude-independent LSTM variant by removing $\mathbf{X}_{\rm mag}$ from the input vector.

\subsection{Bayesian neural network}
\label{sec:BNN}

Bayesian neural networks place prior distributions on the network parameters and infer posterior distributions over the parameters from the data, thereby enabling predictive uncertainties to be estimated within a Bayesian framework \citep[for a review, see][]{BNNReview}. In contrast to standard neural networks, which optimise a single set of deterministic parameters, BNNs treat the network parameters probabilistically.

For a given cluster system, we mapped the projected phase-space distribution of galaxies into a probability density function (PDF). Specifically, the membership-weighted 2D Gaussian kernel density estimate (KDE) was first computed as
\begin{equation}
\tilde{P}(\hat{v}_{\rm los},\rho)
=
\sum_{j=1}^{N_{\rm gal}}
p_{{\rm mem},j}
K_{\mathbf{H}}
\left[
(\hat{v}_{\rm los},\rho)
-
(\hat{v}_{{\rm los},j},\rho_j)
\right].
\label{eq:weighted-KDE}
\end{equation}
Here, $p_{{\rm mem},j}$ is the LSTM-predicted membership probability of galaxy $j$, $K_{\mathbf{H}}$ is a Gaussian kernel with bandwidth matrix $\mathbf{H}$, and $\tilde{P}$ denotes the unnormalised KDE. The coordinates $\hat{v}_{\rm los}$ and $\rho$ are the LOS velocity and projected cluster-centric radius normalised by the corresponding selection limits, as defined above. The KDE was numerically normalised to unit integral over the same finite grid used for the BNN input, yielding the PDF $\hat{P}(\hat{v}_{\rm los},\rho)$, which was used as the BNN input map $\mathbf{x}$.

Applying this procedure to all cluster systems gives a training dataset
$\mathcal{D}=\{(\mathbf{x}_i,y_i)\}_{i=1}^{N}$, where $\mathbf{x}_i$ is the
KDE-based phase-space PDF map and $y_i=\log_{10}\left[M_{200,i}/(h^{-1}\,M_\odot)\right]$
is the corresponding logarithmic cluster mass.

Bayesian inference updates prior beliefs about the network parameters
$\boldsymbol{\theta}$ through Bayes' theorem. Assuming that the targets are conditionally independent given the inputs and network parameters, the likelihood factorises as $p(\mathcal{D}\mid\boldsymbol{\theta})=\prod_{i=1}^{N} p(y_i\mid \mathbf{x}_i,\boldsymbol{\theta})$,
and the posterior distribution is
\begin{equation}
p(\boldsymbol{\theta}\mid \mathcal{D})
=\frac{p(\mathcal{D}\mid \boldsymbol{\theta})\,p(\boldsymbol{\theta})}{p(\mathcal{D})}
=\frac{p(\boldsymbol{\theta})\prod_{i=1}^{N} p(y_i\mid \mathbf{x}_i,\boldsymbol{\theta})}
{\int p(\boldsymbol{\theta}')\prod_{i=1}^{N} p(y_i\mid \mathbf{x}_i,\boldsymbol{\theta}')\,d\boldsymbol{\theta}'}.
\label{eq:Bayes}
\end{equation}

In practice, exact evaluation of this posterior is generally intractable because the evidence term, $p(\mathcal{D})=\int p(\mathcal{D}\mid\boldsymbol{\theta})\,p(\boldsymbol{\theta})\,d\boldsymbol{\theta}$, requires integration over the high-dimensional parameter space of the network. Variational inference addresses this difficulty by introducing a simpler variational distribution $q_{\phi}(\boldsymbol{\theta})$, parametrised by variational parameters $\phi$, to approximate the true posterior. The approximation is obtained by minimising the Kullback--Leibler (KL) divergence,
\begin{equation}
\mathrm{KL}\bigl(q_{\phi}(\boldsymbol{\theta}) \| p(\boldsymbol{\theta}\mid\mathcal{D})\bigr)
= \mathbb{E}_{q_{\phi}}\left[\log\frac{q_{\phi}(\boldsymbol{\theta})}{p(\boldsymbol{\theta}\mid\mathcal{D})}\right].
\label{eq:KL}
\end{equation}

Since $\log p(\mathcal{D}) = \mathrm{ELBO} + \mathrm{KL}\bigl(q_{\phi}(\boldsymbol{\theta}) \| p(\boldsymbol{\theta}\mid\mathcal{D})\bigr)$, minimising the KL divergence is equivalent to maximising the evidence lower bound (ELBO),
\begin{equation}
\mathrm{ELBO}
= \mathbb{E}_{q_{\phi}(\boldsymbol{\theta})}\left[\log p(\mathcal{D}\mid\boldsymbol{\theta})\right]
- \mathrm{KL}\bigl(q_{\phi}(\boldsymbol{\theta})\|p(\boldsymbol{\theta})\bigr).
\label{eq:Elbo}
\end{equation}

After training, predictions for a new input $\mathbf{x}^*$ were obtained by Monte Carlo sampling from the approximate posterior $q_{\phi}(\boldsymbol{\theta})$. The posterior predictive distribution is approximated as
\begin{equation}
p(y^* \mid \mathbf{x}^*, \mathcal{D})
\approx
\frac{1}{n}\sum_{i=1}^{n} p(y^* \mid \mathbf{x}^*, \boldsymbol{\theta}_i),
\qquad
\boldsymbol{\theta}_i \sim q_{\phi}(\boldsymbol{\theta}),
\end{equation}
where $n$ is the number of Monte Carlo samples. For each sampled parameter set $\boldsymbol{\theta}_i$, the network outputs a Gaussian predictive distribution $p(y^* \mid \mathbf{x}^*, \boldsymbol{\theta}_i)$ with mean $\mu_i(\mathbf{x}^*)$ and standard deviation $\sigma_i(\mathbf{x}^*)$. The point prediction is then approximated by the Monte Carlo average of the predictive means,
\begin{equation}
\bar{y}(\mathbf{x}^*)
=
\frac{1}{n}\sum_{i=1}^{n} \mu_i(\mathbf{x}^*).
\label{eq:mean_pred}
\end{equation}
The total predictive variance is estimated as
\begin{equation}
\mathrm{Var}(y^* \mid \mathbf{x}^*, \mathcal{D})
\approx
\frac{1}{n}\sum_{i=1}^{n} \sigma_i^2(\mathbf{x}^*)
+
\frac{1}{n}\sum_{i=1}^{n}\left[\mu_i(\mathbf{x}^*) - \bar{y}(\mathbf{x}^*)\right]^2,
\label{eq:pred_uncertainty}
\end{equation}
where the first term represents the aleatoric uncertainty, and the second term represents the epistemic uncertainty.

\subsection{Model architecture}

The architecture of our model is depicted in Fig.~\ref{fig:model_architecture}. We developed a lightweight LSTM for member identification. The input features were mapped into a higher-dimensional feature space of 128D. The model incorporates a bidirectional LSTM layer, consisting of two stacked LSTM blocks. The bidirectional structure allows the model to use contextual information from both directions along the sequence. The LSTM-extracted features were passed through a linear layer followed by a sigmoid activation, yielding a single output that represents the probability that a galaxy is a cluster member. The LSTM was trained using a binary cross-entropy (BCE) loss, which quantifies the difference between the predicted probabilities and the actual binary labels.

Our model incorporates a BNN to infer cluster masses and
their associated predictive uncertainties. The input to the BNN is the
normalised KDE-based phase-space PDF map $\mathbf{x}$ described in
Sect.~\ref{sec:BNN}. In our fiducial model, this map was constructed using the LSTM-predicted membership probabilities as weights in the KDE, so that galaxies with higher membership probabilities contribute more strongly to the input phase-space distribution.
The BNN uses a hybrid convolutional feature extractor. A deterministic
2D convolutional layer was first applied to the KDE input map,
followed by a 1D convolutional layer, max pooling, dropout, and two deterministic fully connected layers. These layers extract compact
representations of the phase-space PDF for mass inference. Bayesian inference was applied only at the final output stage, which consists of two Bayesian linear layers that predict the mean, $\mu$, and standard deviation, $\sigma$, of a Gaussian likelihood for the cluster mass, respectively. A softplus transformation was applied to the latter to ensure $\sigma>0$. This Bayesian output-layer treatment reduces the computational cost relative to placing priors on all network parameters, while still capturing predictive uncertainty \citep{BnnLastLayer}. The model was trained by minimising the negative ELBO, or equivalently by maximising the ELBO introduced in Sect.~\ref{sec:BNN}.

\subsection{Training methodology}

The LSTM was implemented in PyTorch\footnote{\url{https://pytorch.org}} and
the BNN was developed using Pyro\footnote{\url{https://pyro.ai}} \citep{pyro}, a probabilistic programming library that extends PyTorch. We employed the \texttt{AutoDiagonalNormal} guide for variational inference, which approximates the posterior distribution of the parameters in the two Bayesian output layers with a diagonal Gaussian. In this formulation, each Bayesian parameter is assigned an independent Gaussian variational distribution whose mean and variance are learnt during training. We placed independent standard normal priors, $\mathcal{N}(0,1)$, on the weights and biases of both the mean and standard-deviation output layers. The output of the latter was transformed using a softplus function to ensure a positive predictive standard deviation.

Before training, galaxies were labelled as 1 for members and 0 for interlopers. The input feature matrix was processed using $z$-score normalisation, ensuring that each input feature had zero mean and unit variance. The LSTM and BNN networks were trained with batch sizes of 64 and 100, respectively, and a learning rate of 0.001. Both models were optimised using the Adam optimiser.

During training, we allocated 80\% of the training set to model training and used the remaining 20\% for validation. To avoid overfitting, we implemented early stopping. For the LSTM, training was terminated if the validation loss did not improve for 50 consecutive epochs. For the BNN, which is more computationally demanding, early stopping was applied after 200 consecutive epochs without improvement in the validation loss. In addition, we adopted a learning-rate scheduler to facilitate model convergence. The learning rate was reduced by a factor of 0.1 if the validation loss did not improve for 30 consecutive epochs for the LSTM or 150 consecutive epochs for the BNN.

\section{Results}
\label{sec:result}
\subsection{Member identification}
\label{sec:MI}
\begin{figure}

	\includegraphics[width=\columnwidth]{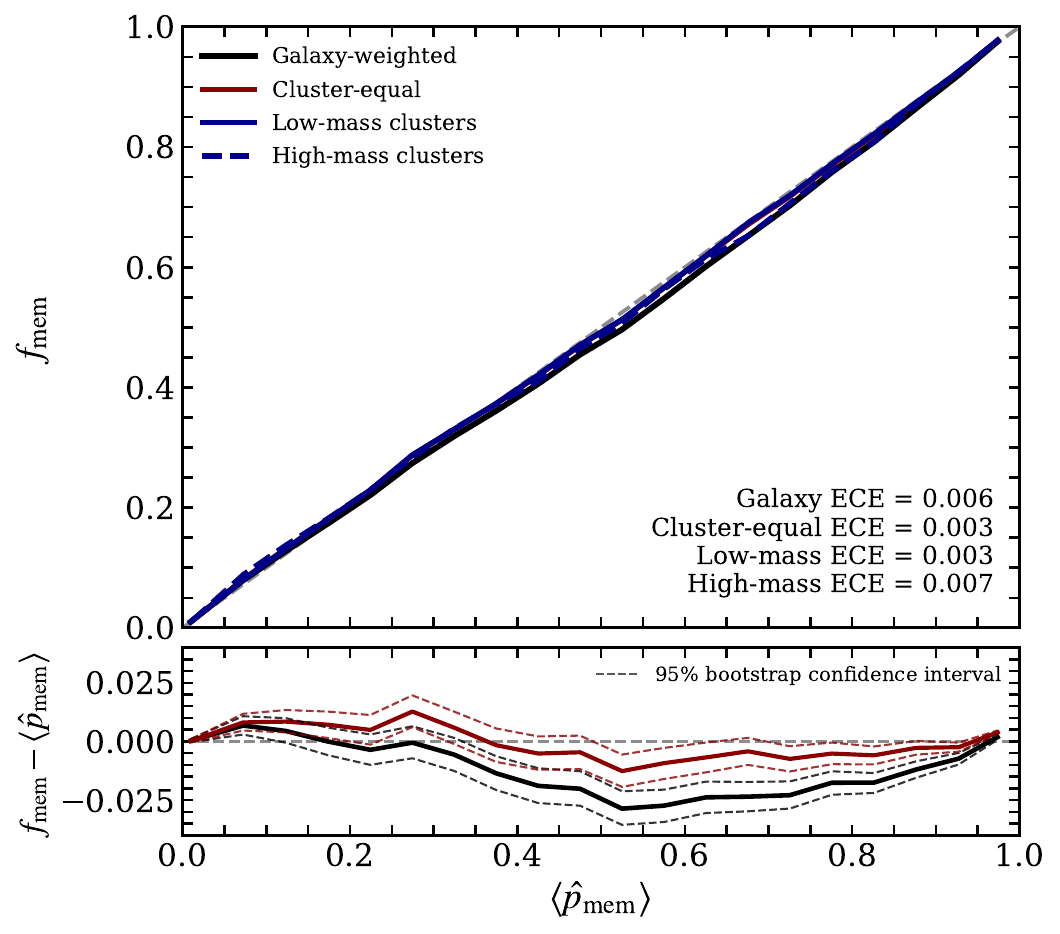}
    \caption{
    Calibration of the predicted membership probability, $p_{\rm mem}$, on the main test set. 
    The member fraction $f_{\rm mem}$ is measured in 20 fixed-width bins of predicted membership probability, $p_{\rm mem}$, with each bin plotted at its weighted mean predicted probability, $\langle p_{\rm mem} \rangle$. The diagonal dashed grey line indicates perfect calibration. The curves compare galaxy-weighted, cluster-equal, and mass-split calibrations, where the mass split is $M_{200}=10^{14.5}\,h^{-1}M_\odot$. Top: ECE values. Bottom: Calibration residual, $\Delta_{\rm cal}=f_{\rm mem}-\langle p_{\rm mem}\rangle$, for the galaxy-weighted and cluster-equal estimators, with dashed curves marking the pointwise 95\% bootstrap confidence limits. The horizontal grey line indicates perfect calibration, $\Delta_{\rm cal}=0$.
    }
    \label{fig:CalibrationCurve}
\end{figure}

To assess the reliability of the predicted membership probabilities, we examined the calibration curve shown in Fig.~\ref{fig:CalibrationCurve}. The calibration curve was constructed by dividing the predicted membership-probability range, $0\leq \hat{p}_{\rm mem}\leq1$, into 20 equal-width bins, corresponding to a bin width of $\Delta p=0.05$ and providing a compromise between resolving probability-dependent calibration variations and limiting sampling noise within individual bins, and comparing the mean predicted probability, $\langle\hat{p}_{\rm mem}\rangle$, with the corresponding member fraction, $f_{\rm mem}$, in each bin. A well-calibrated classifier should follow the one-to-one relation $f_{\rm mem}=\langle p_{\rm mem} \rangle$, shown by the diagonal dashed grey line.

We computed the calibration curve using two weighting schemes. In the galaxy-weighted case, each galaxy has a unit weight, $w_i=1$, so richer clusters contribute more strongly to the final curve. In the cluster-equal case, each galaxy $i\in c$ was assigned $w_i=1/N_c$, where $N_c$ is the number of galaxies in cluster $c$. Thus, the total weight of cluster $c$ is $\sum_{i\in c}w_i=1$, so each cluster contributes equally to the calibration curve. This reduces the impact of cluster richness on the calibration measurement. We also split the sample into low- and high-mass clusters at $M_{200}=10^{14.5}\,h^{-1}M_\odot$ and computed cluster-equal calibration curves for each subsample to test whether the predicted membership probabilities show any mass-dependent calibration bias.

We quantified each calibration curve's deviation from the perfect-calibration relation using the expected calibration error (ECE). The ECE is the sample-weighted average absolute mismatch between the mean predicted membership probability and the observed member fraction across probability bins:
\begin{equation}
{\rm ECE}=\sum_b W_b\left|f_b-\langle p_{\rm mem}\rangle_b\right|,
\label{eq:ece}
\end{equation}
where $b$ indexes the probability bins. For each bin, $f_b$ and $\langle p_{\rm mem}\rangle_b$ are the member fraction and the mean predicted membership probability, computed using the same galaxy weights $w_i$ as the corresponding calibration curve. The bin weight $W_b=\sum_{i\in b}w_i/\sum_i w_i$ is the fraction of the total sample weight contained in bin $b$; therefore, bins containing a larger fraction of the sample contribute more strongly to the ECE. 
We estimated the uncertainty in the calibration residuals using a non-parametric cluster-level bootstrap with 5,000 resamples. In each resample, clusters were drawn with replacement, with all associated LOS realisations and galaxies retained, and the calibration residual, $\Delta_{{\rm cal},b}=f_{{\rm mem},b}-\langle\hat{p}_{\rm mem}\rangle_b$, was recalculated in the original 20 fixed probability bins. The 2.5th and 97.5th percentiles of the resulting bootstrap residuals define the pointwise 95\% confidence limits shown in the lower panel of Fig.~\ref{fig:CalibrationCurve}.

The predicted membership probabilities are well calibrated on the main test set. Both the galaxy-weighted and cluster-equal calibration curves closely follow the one-to-one relation over most of the probability range, with small ECE values of 0.006 and 0.003, respectively. The agreement between these two weighting schemes suggests that the calibration result is not dominated by richer clusters with larger galaxy counts. The lower panel of Fig.~\ref{fig:CalibrationCurve} shows that the calibration residuals remain within $|\Delta_{\rm cal}|\lesssim0.03$ across the probability range, with the largest deviations occurring at intermediate predicted probabilities. The mass-split curves also remain close to the perfect-calibration relation, with ${\rm ECE}=0.003$ for low-mass clusters and ${\rm ECE}=0.007$ for high-mass clusters. Although the high-mass subsample shows a slightly larger calibration error, the deviations remain small in both mass bins, indicating no substantial mass-dependent calibration bias.

\begin{figure}

	\includegraphics[width=\columnwidth]{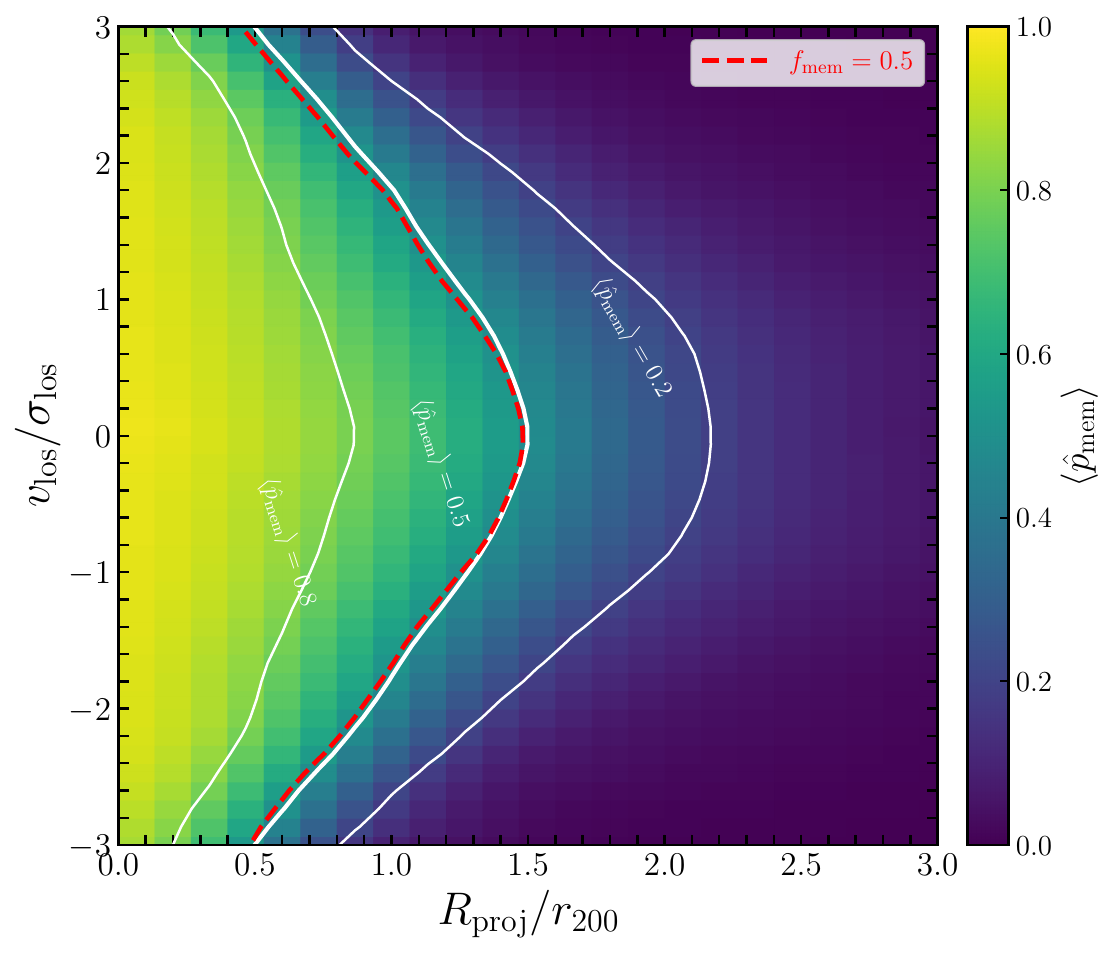}
    \caption{
    Stacked projected phase-space map of the mean LSTM membership probability,
    \(\langle p_{\rm mem} \rangle\), for the main test set. The white contours show \(\langle p_{\rm mem} \rangle = 0.2\), \(0.5\), and \(0.8\),
    while the dashed red contour marks the empirical member-fraction contour
    \(f_{\rm mem}=0.5\).
    }
    \label{fig:Stack-ps}
\end{figure}

In projected phase space, the caustic amplitude \(A(R_{\rm proj})\) defines an approximate radius-dependent envelope in cluster-centric LOS velocity. Galaxies with \(|v_{\rm los}| \lesssim A(R_{\rm proj})\) are preferentially found in the member-dominated region, while the interloper fraction increases beyond this caustic-like envelope \citep{Serra2013}. Since \(p_{\rm mem}\) is a continuous measure of membership likelihood, a physically meaningful prediction should vary coherently across this transition. We therefore examined whether the
isoprobability contours of the LSTM-predicted membership probability field trace a caustic-like probability envelope in the phase-space plane.

To construct this diagnostic, we stacked galaxies in the main test set in the normalised projected phase-space plane, \((R_{\rm proj}/r_{200},v_{\rm los}/\sigma_{\rm los})\), denoted as \((R,v)\). The stacked map was computed over \(0\leq R<4\) and \(-4\leq v<4\), using 30 bins in projected radius and 60 bins in signed LOS velocity. In each phase-space bin, we accumulated the
summed predicted membership probability \(S_p(R,v)=\sum_i p_{{\rm mem},i}\), the summed member galaxy count \(S_y(R,v)=\sum_i y_i\), and the total galaxy count \(N(R,v)\), where \(y_i=1\) for members and \(y_i=0\) for interlopers. To suppress statistical fluctuations, these three stacked maps were smoothed with a Gaussian kernel of width one bin before taking ratios. We thus define
\begin{equation}
\langle p_{\rm mem}\rangle(R,v)=
\frac{S_p^{\rm sm}(R,v)}{N^{\rm sm}(R,v)},
\qquad
f_{\rm mem}(R,v)=
\frac{S_y^{\rm sm}(R,v)}{N^{\rm sm}(R,v)} ,
\end{equation}
where the superscript `sm' denotes the smoothed maps. Bins with \(N_{\rm gal}^{\rm sm}(R,v)<80\) were excluded. The resulting $\langle p_{\rm mem} \rangle$ field is shown in Fig.~\ref{fig:Stack-ps}, together with the $\langle p_{\rm mem} \rangle = 0.8, 0.5$ and \(0.2\) isoprobability contours and the empirical \(f_{\rm mem}=0.5\) contour, obtained by drawing the level set \(f_{\rm mem}(R,v)=0.5\) from the smoothed member-fraction map.

The predicted probability field follows the expected caustic-like pattern. The mean membership probability, $\langle p_{\rm mem} \rangle$, is highest near the cluster centre and at low $|v_{\rm los}|/\sigma_{\rm los}$, before decreasing towards larger projected radii and larger velocity offsets. The three isoprobability contours describe
a probabilistic membership envelope, with the $\langle p_{\rm mem} \rangle = 0.8$ contour enclosing the high-confidence member-dominated region and the $\langle p_{\rm mem} \rangle = 0.2$ contour tracing the lower-probability outskirts. The $\langle p_{\rm mem}\rangle=0.5$ contour provides a probabilistic estimate of the member--interloper transition envelope. Its close agreement with the empirical $f_{\rm mem}=0.5$ contour suggests that the predicted membership probabilities are locally consistent with the member fraction in stacked projected phase space.

\begin{figure}

	\includegraphics[width=\columnwidth]{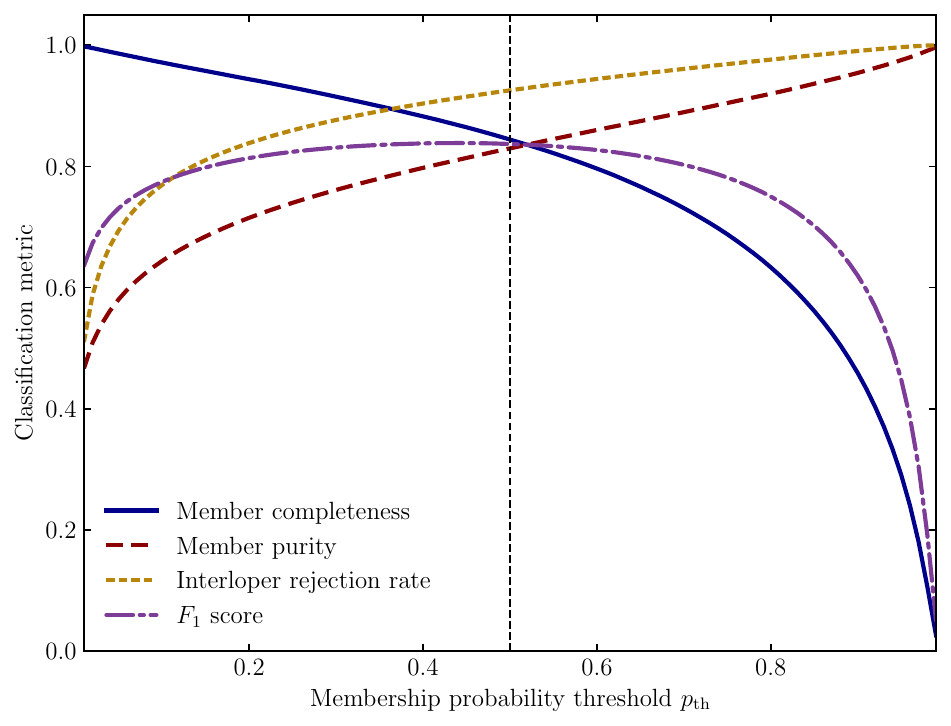}
    \caption{
    Performance of thresholded membership selection as a function of $p_{\rm th}$. Galaxies with $p_{\rm mem}>p_{\rm th}$ were selected as members. The curves show the member completeness, member purity, interloper rejection rate, and $F_1$ score, with the vertical dashed line marking $p_{\rm th}=0.5$.
    }
    \label{fig:Classification-matrix}
\end{figure}

We also examined the performance of a thresholded member selection. For a given probability threshold $p_{\rm th}$, galaxies with $p_{\rm mem}>p_{\rm th}$ were selected as members, while the remaining
galaxies were classified as interlopers.
We define correctly identified members as true positives (TPs) and correctly identified interlopers as true negatives (TNs). Misclassified members are defined as false negatives (FNs), whereas misclassified interlopers are defined as false positives (FPs). We quantified the thresholded selection using the member completeness 
$f_{\rm comp}=\mathrm{TP}/(\mathrm{TP}+\mathrm{FN})$, 
the member purity 
$f_{\rm pur}=\mathrm{TP}/(\mathrm{TP}+\mathrm{FP})$, 
and the interloper rejection rate 
$f_{\rm rej}=\mathrm{TN}/(\mathrm{TN}+\mathrm{FP})$. 
We also report the $F_1$ score, $F_1=2\mathrm{TP}/(2\mathrm{TP}+\mathrm{FP}+\mathrm{FN})$, which summarises the balance between member completeness and member purity.

Figure~\ref{fig:Classification-matrix} shows the performance of the thresholded membership selection as a function of $p_{\rm th}$. We adopted $p_{\rm th}=0.5$ as a natural operating point for the calibrated membership probabilities. As shown in Fig.~\ref{fig:Classification-matrix}, this threshold provides a good balance between member completeness and purity, while the $F_1$ score remains near its maximum over a broad range of thresholds.

\begin{table}
\centering
\caption{Confusion matrix for the thresholded membership selection at $p_{\rm th}=0.5$.}
\label{tab:confusion_matrix}
\begin{tabular}{lcc}
\hline
 & Predicted member & Predicted interloper \\
\hline
True member 
& $84.4\%$ 
& $15.6\%$ \\
True interloper 
& $7.4\%$ 
& $92.6\%$ \\
\hline
\end{tabular}
\tablefoot{
The rows denote the true labels, and the columns denote the predicted labels.
The percentages are normalised by the number of galaxies in each true class.
}
\end{table}

Table~\ref{tab:confusion_matrix} summarises the classification result at $p_{\rm th}=0.5$ using a row-normalised confusion matrix.
The model correctly recovers $84.4\%$ of true members and rejects $92.6\%$ of true interlopers, while $7.4\%$ of interlopers are misclassified as members.
This corresponds to $f_{\rm comp}=84.4\%$, $f_{\rm pur}=83.0\%$, $f_{\rm rej}=92.6\%$, and $F_1=83.7\%$.
We also evaluated the threshold-independent ranking performance using the average precision, ${\rm AP}$, and the area under the receiver operating characteristic curve, ${\rm ROC\mbox{-}AUC}$.
The former summarises the precision--recall performance over all probability thresholds, while the latter measures the overall separability between members and interlopers.
The model achieves ${\rm AP}=0.913$ and ${\rm ROC\mbox{-}AUC}=0.961$, indicating strong threshold-independent ranking performance, with members preferentially assigned high membership probabilities and interlopers mostly assigned lower values.

\begin{table*}[t]
\centering
\caption{
Radial dependence of the thresholded membership selection at $p_{\rm th}=0.5$.
}
\label{tab:membership_radial_bins}

\renewcommand{\arraystretch}{1.65}
\begin{tabular}{lccccc}
\hline\hline
$R_{\rm proj}/r_{200}$ 
& $N_{\rm cluster}$ 
& $f_{\rm mem}$ 
& $f_{\rm comp}$ 
& $f_{\rm pur}$ 
& $f_{\rm rej}$ \\
\hline
$0.0$--$1.0$ 
& $4523$ 
& $0.993^{+0.007}_{-0.041}$ 
& $1.000^{+0.000}_{-0.019}$ 
& $0.998^{+0.002}_{-0.029}$ 
& $0.000^{+0.938}_{-0.000}$ \\

$1.0$--$2.0$ 
& $4523$ 
& $0.885^{+0.098}_{-0.283}$ 
& $0.674^{+0.168}_{-0.200}$ 
& $0.923^{+0.077}_{-0.206}$ 
& $0.500^{+0.457}_{-0.500}$ \\

$2.0$--$3.0$ 
& $4482$ 
& $0.200^{+0.531}_{-0.200}$ 
& $0.000^{+0.100}_{-0.000}$ 
& $0.000^{+0.167}_{-0.000}$ 
& $0.750^{+0.222}_{-0.519}$ \\
\hline
\end{tabular}

\tablefoot{
The statistics were computed in independent bins of $R_{\rm proj}/r_{200}$.
The values denote medians over clusters, with uncertainties indicating the 16th--84th percentile range.
The number of contributing clusters in each radial bin is also listed.
For each metric, cluster--bin values with zero denominators were treated as undefined and excluded from the quoted summary statistics.
}
\end{table*}

Table~\ref{tab:membership_radial_bins} summarises the radial dependence of the thresholded membership selection at $p_{\rm th}=0.5$.
Within $R_{\rm proj}<r_{200}$, the selected sample is nearly complete and highly pure, indicating that the LSTM probabilities retain the dominant virial-region member population.
The broad percentile range of $f_{\rm rej}$ in this bin is mainly caused by the small number of interlopers in many clusters, together with cluster-to-cluster variation in how these few objects are classified.

Beyond $r_{200}$, the performance of the fixed probability threshold changes markedly with projected radius.
At $1<R_{\rm proj}/r_{200}<2$, the selection retains high purity at the expense of completeness.
At $2<R_{\rm proj}/r_{200}<3$, both member completeness and purity decrease substantially, while interloper rejection becomes more effective.
These trends reflect the increasing dominance of interlopers in the projected outskirts and show that a fixed threshold of $p_{\rm th}=0.5$ does not provide uniform classification performance across projected radius.

\subsection{Mass estimation}
\label{sec:massestimation}

\begin{figure*}

	\includegraphics[width=\linewidth]{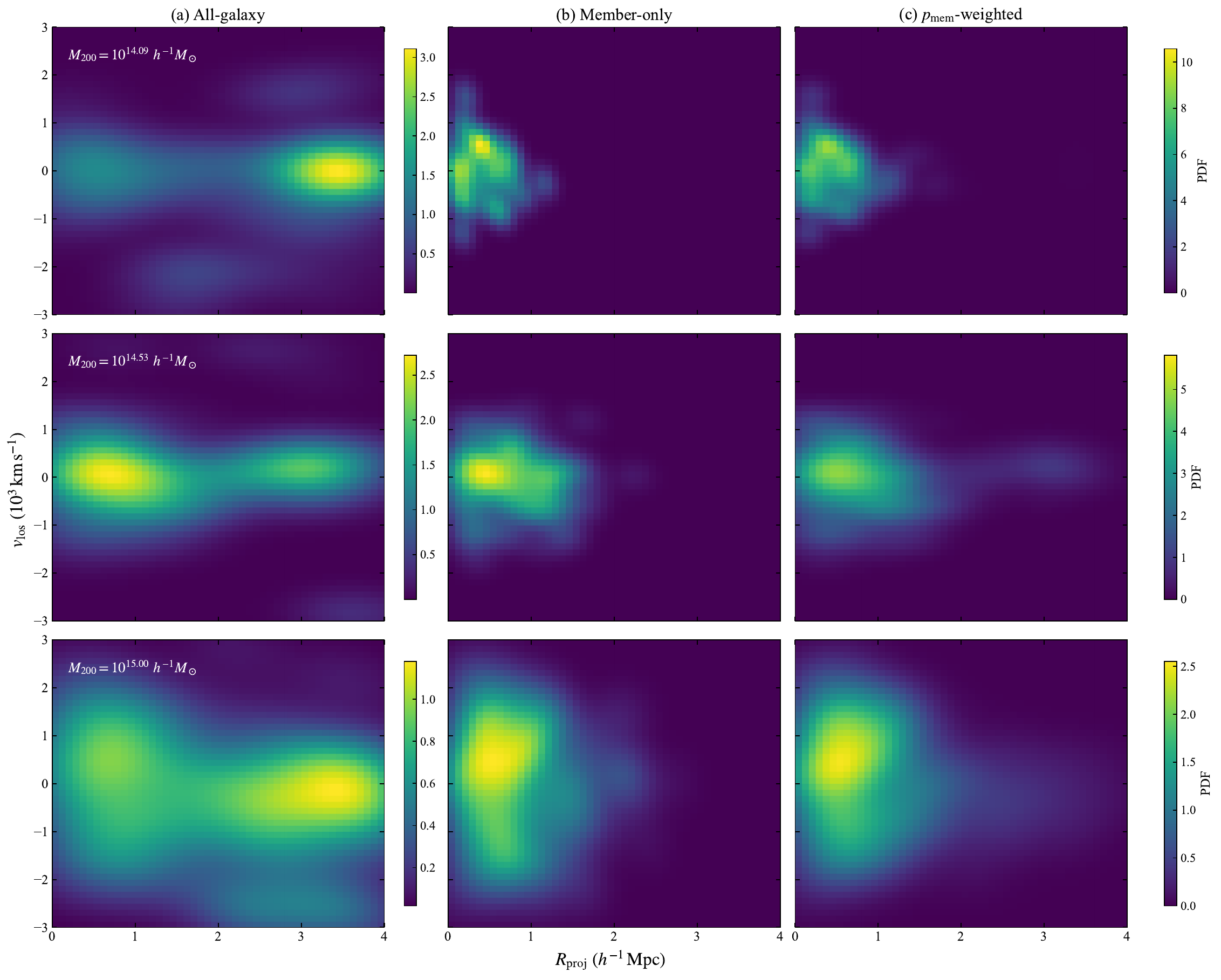}
    \caption{
    Examples of projected phase-space PDFs for three clusters spanning a range of halo masses.
    Columns: PDFs constructed from all galaxies, from members only, and from all galaxies weighted by the predicted membership probability, \(p_{\rm mem}\).
    The \(p_{\rm mem}\)-weighted PDFs in the right column are used as inputs to the BNN mass-estimation model.
    For each row, the member-only and \(p_{\rm mem}\)-weighted panels share the same colour scale.
    }
    \label{fig:PS-pdfs}
\end{figure*}

To evaluate the performance of the BNN for cluster mass estimation, we used the posterior predictive mean as the point estimate. After training, the posterior predictive distribution for each input phase-space PDF was approximated by Monte Carlo sampling from the variational posterior \(q_{\phi}(\boldsymbol{\theta})\). In this work, each prediction was computed from \(N_{\rm MC}=1000\) posterior samples. The point estimate and predictive uncertainty were then computed from these samples using Eqs.~(\ref{eq:mean_pred}) and~(\ref{eq:pred_uncertainty}), respectively.

\begin{figure*}

	\includegraphics[width=\linewidth]{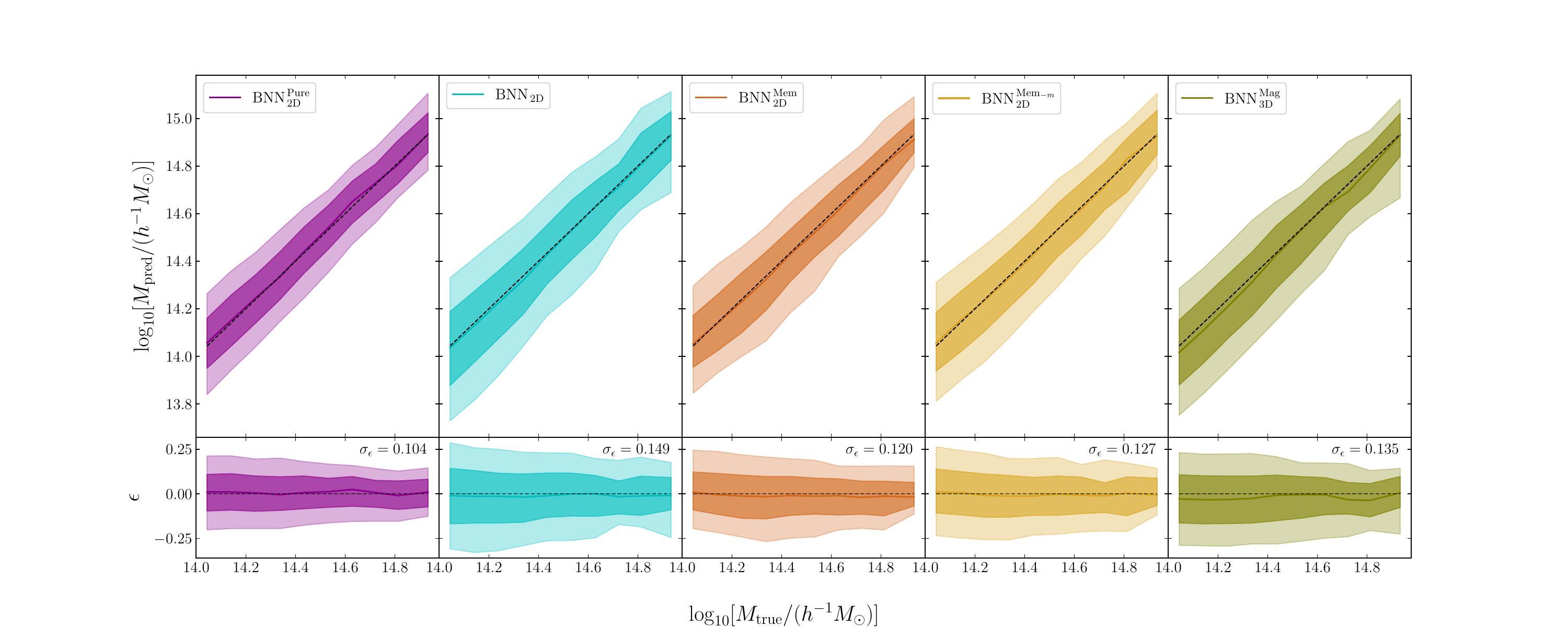}
    \caption{
    Point-estimation performance of the BNN mass-estimation models on the main test set.
    Top panels: Predicted mass as a function of true cluster mass. Bottom panels: Residuals,
    \(\epsilon=\log_{10}(M_{\rm pred}/M_{\rm true})\).
    The dash-dotted black line marks the one-to-one relation, and the solid coloured curves show the median prediction or residual in bins of true mass.
    The shaded regions indicate the central 68\% and 95\% ranges.
    The residual scatter is reported as \(\sigma_\epsilon\) in each panel.
    }
    \label{fig:Mass_Pred}
\end{figure*}

We trained BNNs using several phase-space PDF constructions. 
The baseline \(\mathrm{BNN}_{\mathrm{2D}}\) uses projected phase-space PDFs constructed from galaxy LOS velocities, \(v_{\mathrm{los}}\), and projected radii, \(R_{\mathrm{proj}}\), following the \(\mathrm{CNN}_{\mathrm{2D}}\) architecture of \citet{CNN2D}, with Bayesian inference applied to the final output stage.
We also trained a reference model, \(\mathrm{BNN}_{\mathrm{2D}}^{\mathrm{Pure}}\), using `pure' catalogues that contain only simulation-truth member galaxies. Our fiducial membership-weighted model, \(\mathrm{BNN}_{\mathrm{2D}}^{\mathrm{Mem}}\), uses the same 2D projected phase-space representation, but the input PDF is constructed with each galaxy weighted by the LSTM-predicted membership probability \(p_{\rm mem}\), as in Eq.~(\ref{eq:weighted-KDE}). To isolate the role of magnitude information in the membership-prediction stage, we also trained an ablation model, \(\mathrm{BNN}_{\mathrm{2D}}^{\mathrm{Mem}_{-m}}\). 
This model uses the same membership-weighted PDF construction, but the membership probabilities are predicted by an otherwise identical LSTM network with the magnitude feature removed. 
Here, the subscript \(-m\) denotes the removal of the standardised magnitude feature from the membership model.
Figure~\ref{fig:PS-pdfs} illustrates the construction of the membership-weighted phase-space PDFs for three example clusters spanning a range of halo masses. 
As an additional comparison, we included \(\mathrm{BNN}_{\mathrm{3D}}^{\mathrm{Mag}}\), which incorporates normalised galaxy magnitude directly as a third coordinate in the KDE-based input PDF. 
Unlike the membership-weighted 2D PDF in Eq.~(\ref{eq:weighted-KDE}), where \(p_{\rm mem}\) enters as a galaxy weight, \(\mathrm{BNN}_{\mathrm{3D}}^{\mathrm{Mag}}\) constructs a 3D PDF in \((\hat{v}_{\mathrm{los}}, \rho, \widetilde{M}_r)\), using a kernel \(K_{\mathbf{H}}\left[(\hat{v}_{\mathrm{los}}, \rho, \widetilde{M}_r)-(\hat{v}_{\mathrm{los},j}, \rho_j, \widetilde{M}_{r,j})\right]\).
Figure~\ref{fig:Mass_Pred} summarises the point-estimation performance of these models.
The 2D-based models are broadly unbiased over the full mass range, with median predictions close to the one-to-one relation and median residuals generally centred near zero.
By contrast, \(\mathrm{BNN}_{\mathrm{3D}}^{\mathrm{Mag}}\) tends to underestimate cluster masses, as indicated by its slightly negative median residuals.

The residuals shown in  Fig.~\ref{fig:Mass_Pred} are defined as
\(\epsilon=\log_{10}(M_{\rm pred}/M_{\rm true})\). For all models, the residual scatter decreases towards higher cluster masses, indicating more precise mass estimates in the high-mass regime.
Interloper contamination increases the scatter, as shown by the broader residual distributions of the baseline \(\mathrm{BNN}_{\mathrm{2D}}\) model compared with the idealised \(\mathrm{BNN}_{\mathrm{2D}}^{\mathrm{Pure}}\) reference model.
Relative to the baseline \(\mathrm{BNN}_{\mathrm{2D}}\), the models incorporating additional galaxy information all reduce the residual scatter.
The fiducial membership-weighted model, \(\mathrm{BNN}_{\mathrm{2D}}^{\mathrm{Mem}}\), achieves the lowest scatter among the contaminated-catalogue models, with \(\sigma_\epsilon=0.120\), corresponding to a \(19.5\%\) reduction relative to the baseline \(\mathrm{BNN}_{\mathrm{2D}}\).
The ablation model \(\mathrm{BNN}_{\mathrm{2D}}^{\mathrm{Mem}_{-m}}\) gives \(\sigma_\epsilon=0.127\), a \(14.8\%\) reduction, while the magnitude-augmented model \(\mathrm{BNN}_{\mathrm{3D}}^{\mathrm{Mag}}\) gives \(\sigma_\epsilon=0.135\), a \(9.4\%\) reduction.
Thus, using membership probabilities to weight the projected phase-space PDFs provides the strongest improvement in point-estimation precision, with the inclusion of magnitude information in the membership-prediction stage giving a modest additional gain.

\begin{figure}

	\includegraphics[width=\columnwidth]{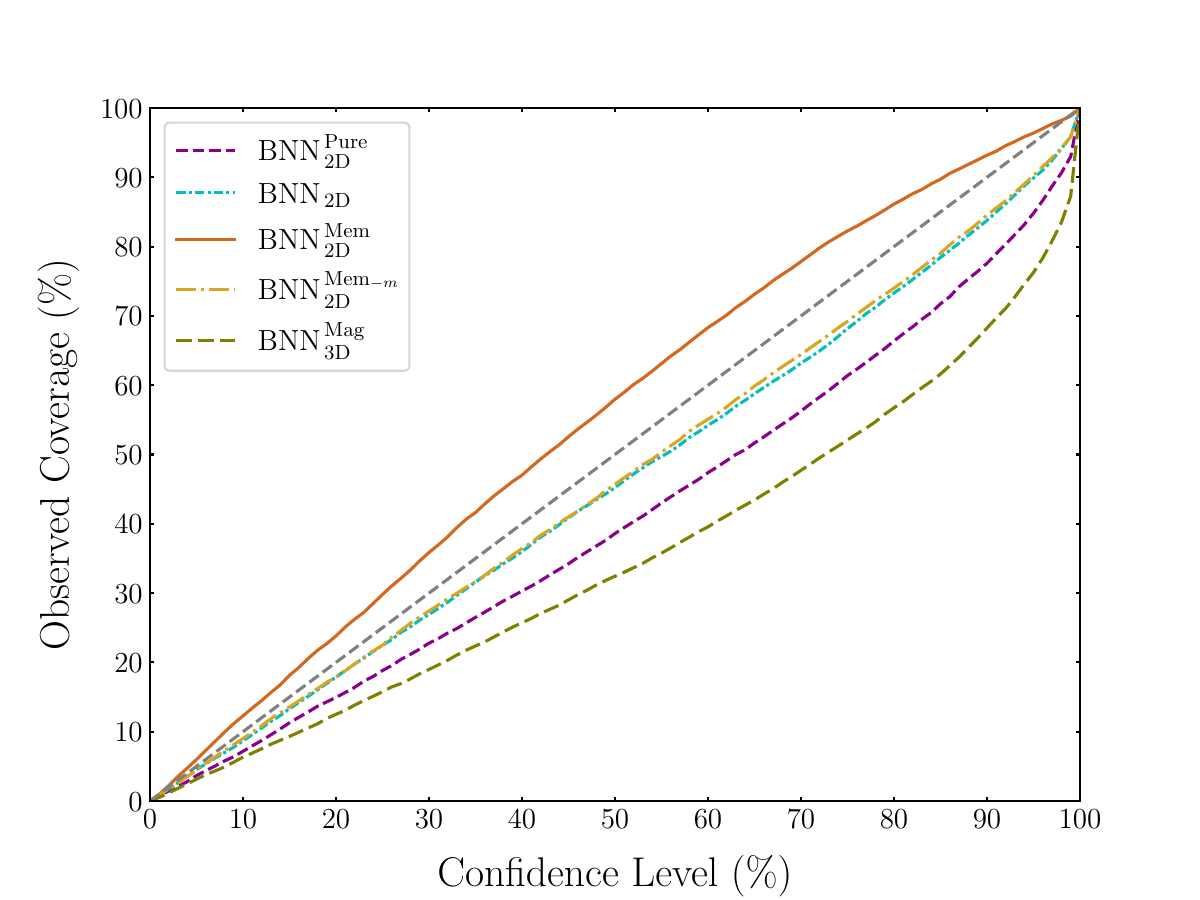}
    \caption{
    Calibration curves for the predictive uncertainties of different cluster mass-estimation models on the main test set. The horizontal axis shows the nominal confidence level of the central predictive interval, and the vertical axis shows the observed coverage, defined as the fraction of test clusters whose true masses fall within that interval. The diagonal dashed line indicates ideal calibration.
    }
    \label{fig:ConfidenceLevel}
\end{figure}

\begin{table}
\caption{Effect of galaxy sequence ordering on membership identification and cluster mass estimation in the main test set.}
\label{tab:sequence_ordering}
\centering
\renewcommand{\arraystretch}{1.25}
\begin{tabular}{lccc}
\hline\hline
Ordering & AP & ECE & $\sigma_\epsilon$ (dex) \\
\hline
Nearest-neighbour & 0.913 & 0.003 & 0.120 \\
Random & 0.911 & 0.009 & 0.124 \\
\hline
\end{tabular}

\tablefoot{
The ECE was computed using the cluster-equal weighting defined in Sect.~\ref{sec:MI}.
}
\end{table}

To assess the effect of sequence construction, we compared the fiducial ordering with the random ordering described in Sect.~\ref{sec:lstm}. We reconstructed the membership-weighted KDE inputs from the random-order LSTM predictions and retrained the corresponding \(\mathrm{BNN}_{\mathrm{2D}}^{\mathrm{Mem}}\). As shown in Table~\ref{tab:sequence_ordering}, the two schemes yield comparable overall performance in membership identification and mass estimation, with the fiducial ordering giving a lower calibration error and a slightly smaller mass residual scatter. Under the same training settings and stopping criterion, the random-order LSTM was trained for approximately 300 more epochs than the nearest-neighbour model. This suggests that the adopted ordering may facilitate optimisation.

Figure~\ref{fig:ConfidenceLevel} shows the calibration curves for the predictive uncertainties of the different mass-estimation models. The curves compare the nominal confidence levels of the central predictive intervals with their empirical coverage fractions. Here, the intervals are constructed from the posterior predictive mean and the total predictive standard deviation using a Gaussian approximation to the posterior predictive distribution.
The different mass-estimation models show distinct uncertainty-calibration behaviours.
The baseline $\mathrm{BNN}_{\mathrm{2D}}$ and the magnitude-independent membership model $\mathrm{BNN}_{\mathrm{2D}}^{\mathrm{Mem}_{-m}}$ remain relatively close to the ideal relation, although both show moderate under-coverage.
The idealised $\mathrm{BNN}_{\mathrm{2D}}^{\mathrm{Pure}}$ model exhibits stronger under-coverage, while $\mathrm{BNN}_{\mathrm{3D}}^{\mathrm{Mag}}$ shows the largest deviation from ideal calibration.
In contrast, the fiducial $\mathrm{BNN}_{\mathrm{2D}}^{\mathrm{Mem}}$ model systematically over-covers, indicating more conservative predictive intervals.
These results show that improved point-estimation precision does not necessarily imply better uncertainty calibration.

\begin{table*}
\centering
\caption{
Summary statistics of the uncertainty-normalised residuals.
}
\label{tab:normalised_residuals}
\renewcommand{\arraystretch}{1.25}
\begin{tabular}{lccccccc}
\hline
Model
& \(\langle\tilde{\epsilon}\rangle\)
& \(\sigma_{\tilde{\epsilon}}\)
& Skewness
& Excess kurtosis
& \(P(|\tilde{\epsilon}|<1)\)
& \(P(|\tilde{\epsilon}|<2)\)
& \(P(|\tilde{\epsilon}|<3)\) \\
\hline
\(\mathrm{BNN}_{\mathrm{2D}}^{\mathrm{Pure}}\)
& 0.144 & 1.409 & 0.312 & 0.858 & 54.6\% & 85.7\% & 96.0\% \\
\(\mathrm{BNN}_{\mathrm{2D}}\)
& $-0.042$ & 1.240 & 0.375 & 1.689 & 61.6\% & 90.4\% & 97.6\% \\
\(\mathrm{BNN}_{\mathrm{2D}}^{\mathrm{Mem}}\)
& 0.039 & 0.922 & 0.840 & 3.599 & 76.2\% & 96.7\% & 99.4\% \\
\(\mathrm{BNN}_{\mathrm{2D}}^{\mathrm{Mem}_{-m}}\)
& 0.098 & 1.236 & 0.672 & 2.140 & 62.9\% & 90.8\% & 97.7\% \\
\(\mathrm{BNN}_{\mathrm{3D}}^{\mathrm{Mag}}\)
& $-0.267$ & 1.783 & 0.419 & 4.013 & 46.2\% & 77.3\% & 91.9\% \\
\hline
\end{tabular}

\tablefoot{
Here, \(\tilde{\epsilon}=\epsilon/\sigma_{\rm pred}\), where
\(\epsilon=\log_{10}(M_{\rm pred}/M_{\rm true})\).
For a well-calibrated Gaussian predictive distribution,
\(\langle\tilde{\epsilon}\rangle\simeq0\),
\(\sigma_{\tilde{\epsilon}}\simeq1\), and both the skewness and excess kurtosis are close to zero.
}
\end{table*}

To further characterise the predicted uncertainty scale, we examined the uncertainty-normalised residuals,
\(\tilde{\epsilon}=\epsilon/\sigma_{\rm pred}\).
For an unbiased and well-calibrated Gaussian predictive distribution, \(\tilde{\epsilon}\) should have a mean close to zero and a standard deviation close to unity.
The normalised-residual statistics are consistent with the coverage behaviour described above, while also revealing departures from Gaussianity.
In particular, the fiducial $\mathrm{BNN}_{\mathrm{2D}}^{\mathrm{Mem}}$ model has a narrower normalised-residual distribution, consistent with its conservative predictive intervals, whereas $\mathrm{BNN}_{\mathrm{3D}}^{\mathrm{Mag}}$ shows the broadest distribution and a non-zero mean residual.
The non-zero skewness and excess kurtosis of several models further indicate that the normalised residual distributions are not strictly Gaussian, suggesting that the Gaussian approximation does not fully characterise the predictive uncertainties.

\subsection{Effects of miscentring and data incompleteness}

\begin{figure*}

	\includegraphics[width=\linewidth]{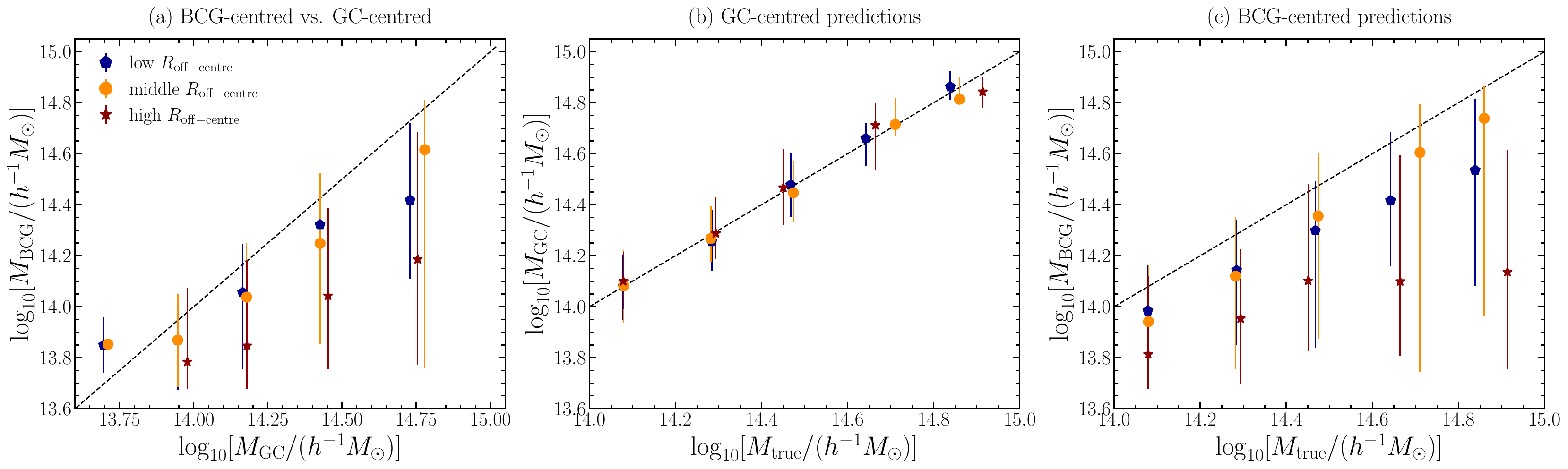}
    \caption{
    Mass comparisons for the off-centre test set, where \(M_{\mathrm{GC}}\) and \(M_{\mathrm{BCG}}\) denote the mass predictions obtained from projections centred on the gravitational centre and on the BCG, respectively, using \(\mathrm{BNN}_{\mathrm{\,2D}}^{\mathrm{\,Mem}}\).
    The sample is divided into three equal-number bins according to \(R_{\rm off-centre}\), defined as the projected separation between the gravitational centre and the BCG centre, with bin boundaries at \(0.271\) and \(0.735\,h^{-1}\,\mathrm{Mpc}\).
    The points indicate medians; the error bars indicate the 16th--84th percentile ranges; and dashed lines mark the one-to-one relations.
}
    \label{fig:Miscentering}
\end{figure*}

In this subsection, we discuss employing the $\mathrm{BNN}_{\mathrm{\,2D}}^{\mathrm{\,Mem}}$ model to assess the effects of centre offset and galaxy incompleteness on cluster mass estimation.
We assessed the sensitivity of the network to cluster miscentring using the mock off-centre test set described in Sect.~\ref{sec:Dataset}. For each system, we compared predictions obtained from projections centred on the gravitational centre with those obtained after recentring the projection on the BCG. In relaxed systems, the BCG is generally expected to lie close to the gravitational centre, whereas significant offsets between the BCG or optical centre and other cluster-centre definitions are commonly associated with dynamically disturbed systems \citep{Offcentre-Lopes,Offcentre-Seppi}. We denote the projected separation between the two adopted centres as \(R_{\rm off-centre}\).

We divided the off-centre test set into three equal-number bins according to \(R_{\rm off-centre}\), with bin boundaries at \(0.271\) and \(0.735\,h^{-1}\,\mathrm{Mpc}\). We denote the cluster-mass predictions obtained from projections centred on the gravitational centre and on the BCG as $M_{\mathrm{GC}}$ and $M_{\mathrm{BCG}}$, respectively. Figure~\ref{fig:Miscentering} compares these predictions with the true cluster masses for the three $R_{\rm off-centre}$ bins. The \(M_{\rm GC}\)--\(M_{\rm true}\) relation remains similar to that found for the main test set, with no strong systematic separation among the three \(R_{\rm off-centre}\) bins. The corresponding residual scatters are $\sigma_\epsilon=0.113$, $0.124$, and $0.123$ for the low-, intermediate-, and high-offset bins, respectively. These values are comparable to $\sigma_\epsilon=0.120$ obtained for the main test set and show no clear monotonic trend with $R_{\rm off-centre}$.
Thus, we find no clear degradation of the gravitational-centre-based predictions with increasing $R_{\rm off-centre}$.
By contrast, systems in the largest \(R_{\rm off-centre}\) bin tend to have lower \(M_{\rm BCG}\) at fixed \(M_{\rm true}\), particularly at the high-mass end. This behaviour is also seen in the \(M_{\rm GC}\)--\(M_{\rm BCG}\) comparison, where the largest-offset systems more often lie below the one-to-one relation, indicating \(M_{\rm GC}>M_{\rm BCG}\). At the high-mass end, the small number of systems within each \(R_{\rm off-centre}\) bin leads to broader percentile ranges. Quantitatively, relative to the predictions obtained from projections centred on the gravitational centre, the median mass changes are $-16.8\%$, $-26.0\%$, and $-47.6\%$ for the low-, intermediate-, and high-offset subsamples, respectively.

Overall, the predictions centred on the gravitational centre are relatively insensitive to $R_{\rm off-centre}$, while the BCG-centred predictions show larger deviations for systems with the largest projected separations. If $R_{\rm off-centre}$ is interpreted as an indirect indicator of dynamical state, this contrast suggests that, for the present model and sample, the inferred masses are more sensitive to the choice of centre than to dynamical state. However, because $R_{\rm off-centre}$ also directly quantifies BCG miscentring, the present comparison cannot fully disentangle the effects of dynamical state and centring.

\begin{figure*}
    \includegraphics[width=\linewidth]{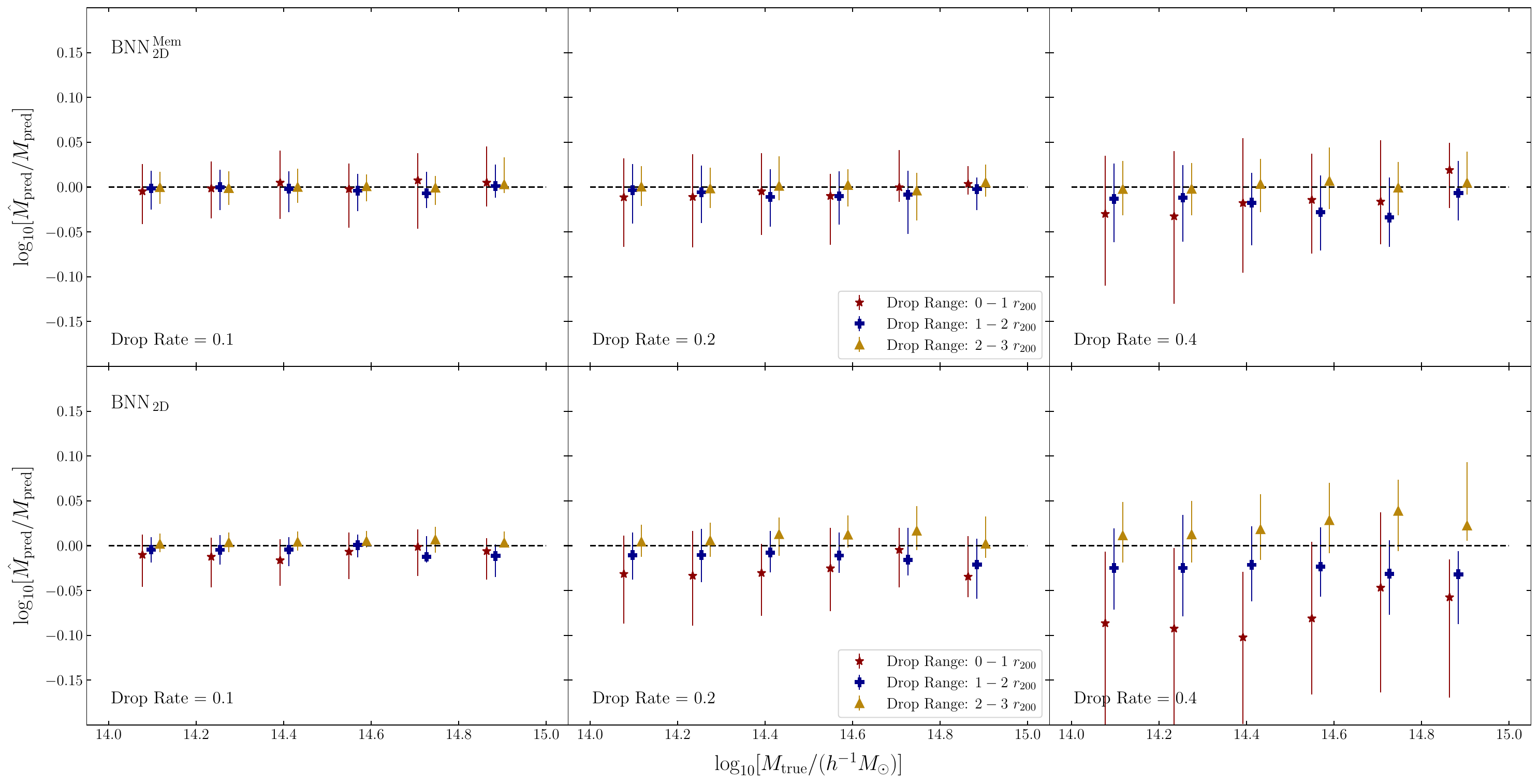}
    \caption{
    Effect of radially dependent incompleteness on the BNN mass predictions.
   Top: Membership-weighted model, \(\mathrm{BNN}_{\mathrm{2D}}^{\mathrm{Mem}}\). Bottom: Phase-space-only baseline, \(\mathrm{BNN}_{\mathrm{2D}}\).
    The horizontal axis shows the true cluster mass, and the vertical axis shows
    \(\log_{10}(\hat{M}_{\rm pred}/M_{\rm pred})\),
    where \(M_{\rm pred}\) and \(\hat{M}_{\rm pred}\) are the predictions from the original and incomplete catalogues, respectively.
    Columns: Galaxy missing fractions of 0.1, 0.2, and 0.4.
    The different colours and marker shapes denote the radial ranges from which galaxies are removed:
    \(0\)--\(1\,r_{200}\), \(1\)--\(2\,r_{200}\), and \(2\)--\(3\,r_{200}\).
    The symbols and error bars show the median and central 68\% interval in each true-mass bin.
    }
    \label{fig:missingdata}
\end{figure*}

In addition to off-centring, we also tested the impact of incomplete galaxy information on the mass estimates. We randomly selected 1,000 clusters from the main test set and artificially removed galaxies from different projected radial ranges at missing fractions of 0.1, 0.2, and 0.4. Figure~\ref{fig:missingdata} shows the prediction shift induced by incompleteness, measured as \(\log_{10}(\hat{M}_{\rm pred}/M_{\rm pred})\), where \(M_{\rm pred}\) and \(\hat{M}_{\rm pred}\) are the predictions from the original and incomplete catalogues, respectively. For the membership-weighted model, \(\mathrm{BNN}_{\mathrm{2D}}^{\mathrm{Mem}}\), the prediction shifts remain small across the tested radial ranges at missing fractions of 0.1 and 0.2. This suggests that incompleteness at these levels does not strongly perturb the membership-weighted mass estimation. With increasing incompleteness, however, the model begins to show a tendency towards lower predicted masses relative to the
original catalogues. This effect is strongest when galaxies are removed
from the inner region, indicating that the mass prediction is particularly
sensitive to incompleteness within $r_{200}$.
For the case with the strongest incompleteness in the central region, in which 40\% of the
eligible non-central galaxies within $r_{200}$ are removed, the median
inferred mass decreases by 6.1\% relative to the complete catalogues.

For comparison, the phase-space-only baseline \(\mathrm{BNN}_{\mathrm{2D}}\) shows a stronger radius-dependent response under severe incompleteness. At a missing fraction of 0.4, removing inner galaxies also lowers the predicted mass, while removing galaxies from the outer region can shift the prediction upwards relative to the original catalogue. This opposite response may be related to the normalisation of the input phase-space PDF, because removing outer galaxies reduces the relative contribution of the outskirts and can make the remaining normalised PDF appear more centrally concentrated. The weaker response of \(\mathrm{BNN}_{\mathrm{2D}}^{\mathrm{Mem}}\) indicates that membership weighting reduces, but does not eliminate, the impact of radially dependent incompleteness.

\subsection{Application to observed clusters}
\label{sec:application}

To examine the behaviour of our model on observed systems, we performed an external consistency check using a sample of observed galaxy clusters. We first defined an observational parent pool based on the cluster compilation of \citet{GCListGroener}. From this catalogue, we selected systems with published $M_{200}$ estimates in the range $10^{14}$--$10^{15}\,M_\odot$, SDSS Data Release 17 (DR17) spectroscopic coverage \citep{SDSSDR17}, and redshifts $z\leq0.3$. We also required at least ten galaxies with spectroscopic redshifts within a BCG-centred aperture of $R_{\rm proj}\leq0.5\,h^{-1}\,\mathrm{Mpc}$, ensuring a minimum level of central spectroscopic sampling. This selection yields a parent pool of 87 clusters.

From this parent pool, we constructed a test sample of 16 clusters, comprising eight deliberately selected systems spanning a range of astrophysical environments and dynamical states and eight randomly drawn clusters. The deliberately selected subsample includes four well-studied systems---Perseus, Coma, Abell~2142, and Abell~2199---the poorer system MKW~4 \citep{OSullivan2003MKW4}, the evolved off-axis merger Abell~119 \citep{Watson2023A119}, the unrelaxed merging cluster Abell~1132 \citep{Wilber2018A1132}, and Abell~2061, which shows merger-related disturbances and forms part of the pre-merger Abell~2061--Abell~2067 system \citep{Marini2004A2061,Pignataro2024A2061}.
For each cluster, we centred the system on the BCG and selected galaxies within a projected aperture of $R_{\rm proj}\leq4\,h^{-1}\,\mathrm{Mpc}$ and a LOS velocity window of $|v_{\rm los}|\leq3000~\mathrm{km\,s^{-1}}$.

We then adopted the reported $M_{200}$ values from recent observational studies as the literature masses, $M_{\rm lit}$, for each system. The measurements were compiled from weak-lensing, X-ray, or dynamical analyses. We note that the resulting \(M_{\rm lit}\) values are not derived from a single homogeneous analysis pipeline and therefore may include method-dependent systematic differences. Predicted masses, \(M_{\rm pred}\), were obtained using the same inference framework as in the simulations. For the SDSS spectroscopic application, we adopted the previously introduced \(\mathrm{BNN}_{\mathrm{2D}}^{\mathrm{Mem}_{-m}}\) model, which excludes magnitude information from the LSTM membership-model inputs, thereby avoiding possible biases due to differences between the simulated CSST \(r\)-band magnitudes and the SDSS photometric system.

\begin{table*}
\caption{Dynamical mass inference for observed clusters.}
\label{tab:inference_real_clusters}
\centering
\begingroup
\renewcommand{\arraystretch}{1.35}
\setlength{\tabcolsep}{4pt}
\begin{tabular}{lcccc}
\hline\hline
Cluster & \(z\) & \(M_{\rm pred}\) & \(M_{\rm lit}\) & Lit. source \\
\hline
Perseus          & 0.0170 & \(15.018 \pm 0.060\) & \(14.909^{+0.063}_{-0.074}\) & Dyn. (1) \\
Coma             & 0.0230 & \(14.778 \pm 0.078\) & \(14.862^{+0.045}_{-0.053}\) & Dyn. (2) \\
Abell 2199       & 0.0300 & \(14.703 \pm 0.083\) & \(14.923^{+0.070}_{-0.048}\) & X-ray (3) \\
Abell 2142       & 0.0909 & \(14.925 \pm 0.064\) & \(14.965^{+0.016}_{-0.016}\) & X-ray (4) \\
MKW 4            & 0.0200 & \(14.133 \pm 0.095\) & \(13.817^{+0.062}_{-0.073}\) & X-ray (5) \\
Abell 2061       & 0.0783 & \(14.237 \pm 0.137\) & \(14.509\) & LOSVD (6) \\
Abell 119        & 0.0446 & \(14.720 \pm 0.083\) & \(14.722\) & CHANCES/MENeaCS (7) \\
Abell 1132       & 0.1369 & \(14.360 \pm 0.088\) & \(14.421^{+0.004}_{-0.004}\) & Caustic (8) \\
Abell 1066       & 0.0689 & \(14.508 \pm 0.067\) & \(14.710\) & Dyn. (9) \\
Abell 1291A      & 0.0513 & \(13.748 \pm 0.132\) & \(13.830\) & Dyn. (9) \\
Abell 2428       & 0.0840 & \(13.911 \pm 0.130\) & \(14.027^{+0.012}_{-0.013}\) & Caustic (8) \\
RXCJ1210.3+0523  & 0.0760 & \(14.217 \pm 0.107\) & \(13.860\) & Caustic (10) \\
Abell 1773       & 0.0774 & \(14.478 \pm 0.072\) & \(14.810\) & Dyn. (9) \\
Abell 2064       & 0.0737 & \(14.176 \pm 0.082\) & \(14.460\) & Dyn. (9) \\
Abell 2670       & 0.0760 & \(14.525 \pm 0.111\) & \(14.608^{+0.009}_{-0.009}\) & Caustic (8) \\
ZwCl0104.4+0048  & 0.2540 & \(13.925 \pm 0.112\) & \(14.367^{+0.159}_{-0.188}\) & WL (11) \\
\hline
\end{tabular}
\endgroup

\tablefoot{
All masses are \(\log_{10}[M_{200}/(h^{-1} M_\odot)]\).
Uncertainties in \(M_{\rm pred}\) are \(1\sigma\) predictive uncertainties from the
\(\mathrm{BNN}_{\mathrm{2D}}^{\mathrm{Mem}_{-m}}\).
Literature methods are abbreviated as X-ray, Dyn. (dynamical/kinematic),
LOSVD (line-of-sight velocity dispersion), Caustic (caustic mass estimate),
and WL (weak lensing).
CHANCES (CHileAN Cluster galaxy Evolution Survey)/MENeaCS
(Multi-Epoch Nearby Cluster Survey) denotes a homogeneous catalogue mass
estimate based on the MENeaCS input catalogue.
We adopt \(h=0.6766\), consistent with the JT1G cosmology.
}

\tablebib{
(1)~\citet{Aguerri2020Perseus};
(2)~\citet{Pedratti2026ComaDESI};
(3)~\citet{Babyk2014};
(4)~\citet{XrayMassprofile};
(5)~\citet{Sarkar2021MKW4};
(6)~\citet{GCListGroener};
(7)~\citet{Sifon2025CHANCES};
(8)~\citet{Pizzardo2021};
(9)~\citet{Kopylova2022};
(10)~\citet{Smith2023};
(11)~\citet{Okabe2016}.
}
\end{table*}

\begin{figure}
    \centering
    \includegraphics[width=\columnwidth]{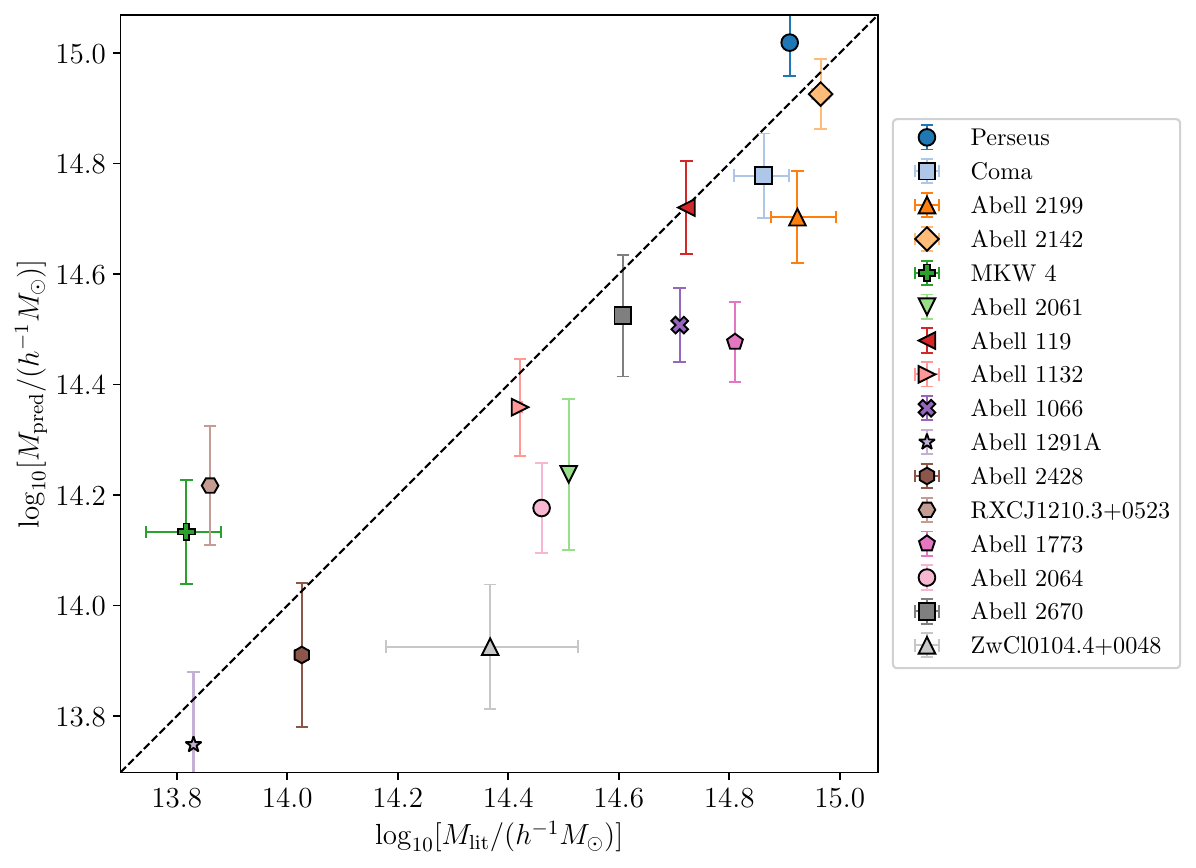}
    \caption{
    Comparison between the predicted cluster masses \(M_{\rm pred}\) from the \(\mathrm{BNN}_{\mathrm{2D}}^{\mathrm{Mem}_{-m}}\) model and literature masses \(M_{\rm lit}\) for the observational sample. Literature masses are reported as \(M_{200}\). The dashed line denotes the one-to-one relation.
    }

    \label{fig:lit_comp}
\end{figure}

Figure~\ref{fig:lit_comp} compares the model-predicted masses with the adopted literature masses for the 16 observed clusters listed in Table~\ref{tab:inference_real_clusters}.
The residuals, defined as $\Delta\log M=\log M_{\rm pred}-\log M_{\rm lit}$, have a mean offset of $-0.090\pm0.054$ dex and a sample standard deviation of $0.216$ dex, where the uncertainty on the mean denotes the standard error. The negative mean offset suggests a modest tendency for the model to predict lower masses than the literature estimates; however, it remains consistent with zero at the $2\sigma$ level within this limited sample.
We further compared systems with literature masses derived from conventional galaxy-dynamical methods (Dyn., line-of-sight velocity dispersion, or Caustic) with the remaining systems.
For the eleven systems in the dynamical subset, the mean offset is $\langle\Delta\log M\rangle_{\rm dyn}=-0.095\pm0.059$ dex, with a sample standard deviation of $0.196$ dex.
For the remaining five systems, whose literature estimates are listed as X-ray, the CHileAN Cluster galaxy Evolution Survey (CHANCES)/Multi-Epoch Nearby Cluster Survey (MENeaCS) catalogue, or weak lensing, the mean offset is $-0.078\pm0.126$ dex, with a sample standard deviation of $0.281$ dex.
The two subsets have similar mean offsets, with a larger residual scatter in the remaining five systems.

\section{Conclusion}
\label{sec:conclusion}

We presented an LSTM-based framework for assigning membership probabilities to galaxies within a fixed projected aperture of $R_{\rm proj}=4\,h^{-1}\mathrm{Mpc}$ around galaxy clusters.
On the main test set, the predicted membership probabilities are well calibrated and show no substantial mass-dependent calibration bias.
In projected phase space, these probabilities form a caustic-like membership envelope whose $\langle p_{\rm mem}\rangle=0.5$ contour broadly traces the empirical member--interloper transition.

As a binary member-classification benchmark, adopting $p_{\rm th}=0.5$ yields a good balance between completeness and purity on the main test set, with $f_{\rm comp}=84.4\%$, $f_{\rm pur}=83.0\%$, and $F_1=83.7\%$.
The model also shows strong threshold-independent ranking performance, with ${\rm AP}=0.913$ and ${\rm ROC\mbox{-}AUC}=0.961$.
The $p_{\rm th}=0.5$ selection is nearly complete and highly pure within $R_{\rm proj}<r_{200}$, but its performance varies substantially at larger projected radii, showing that a single global threshold does not provide a uniform balance between completeness and purity across projected radius.

The order sensitivity of recurrent architectures makes the construction of the input pseudo-sequence an important part of the LSTM design.
Our adopted ordering is motivated by the local coherence of cluster galaxies in projected phase space.
By iteratively connecting galaxies with similar projected positions and LOS velocities, the pseudo-sequence is intended to facilitate the use of local contextual information by the recurrent network.
Nevertheless, this ordering should not be interpreted as a physical temporal sequence or as a reconstruction of the galaxies' accretion histories.
The comparison with random ordering shows similar overall performance in membership identification and mass estimation. This may partly reflect the fact that randomising the sequence changes the order and adjacency of the galaxies but leaves the feature values associated with each galaxy unchanged. In particular, the projected positions and normalised LOS velocities used to construct the fiducial ordering remain available to the model, together with local-density and cluster-level information. The bidirectional LSTM can still use contextual information from both directions along the sequence. We retain the nearest-neighbour ordering as a consistent procedure for constructing input sequences from simulated and observed galaxy catalogues. Whether other ordering schemes can further improve performance remains to be explored.

For cluster mass inference, we used the LSTM membership probabilities to construct membership-weighted projected phase-space PDFs as inputs to the BNN mass-estimation model.
Among the contaminated-catalogue models, the fiducial \(\mathrm{BNN}_{\mathrm{2D}}^{\mathrm{Mem}}\) model achieves the lowest residual scatter, \(\sigma_\epsilon=0.120\), corresponding to a \(19.5\%\) reduction relative to the unweighted \(\mathrm{BNN}_{\mathrm{2D}}\) baseline.
The magnitude-independent membership model, $\mathrm{BNN}_{\mathrm{2D}}^{\mathrm{Mem}_{-m}}$, yields a smaller improvement, while directly incorporating magnitude as a third KDE coordinate in $\mathrm{BNN}_{\mathrm{3D}}^{\mathrm{Mag}}$ is less effective than using it to refine the LSTM membership probabilities.

The predictive-uncertainty calibration varies across the mass-estimation models.
The fiducial $\mathrm{BNN}_{\mathrm{2D}}^{\mathrm{Mem}}$ model achieves the best point-estimation precision among the contaminated-catalogue models but produces comparatively conservative predictive intervals, whereas the other models show varying degrees of under-coverage.
The normalised-residual distributions also exhibit departures from Gaussianity, showing that improved mass precision does not necessarily imply well-calibrated predictive uncertainties.
These departures from ideal uncertainty calibration may partly reflect the approximate Bayesian treatment adopted here, in which only the final output stage is treated probabilistically while the convolutional feature extractor remains deterministic, together with the limited flexibility of the mean-field variational posterior \citep{watson2021latent,harrison2024variational,foong2019inbetween}.

We further tested the model under two observationally motivated systematics: cluster miscentring and radially dependent incompleteness.
For systems with large BCG--gravitational-centre offsets, BCG-centred predictions are typically lower than their gravitational-centre counterparts, showing that large centre offsets can affect the inferred masses.
Radially dependent incompleteness produces increasingly large prediction shifts as the missing fraction increases, with the strongest effect occurring when galaxies are removed from within $r_{200}$. Both miscentring and incompleteness may also affect the inferred masses indirectly by altering the nearest-neighbour pseudo-sequence, making the sequence construction an additional channel through which observational systematics can propagate into the LSTM membership probabilities.

As an initial test on observed systems, we applied the magnitude-independent $\mathrm{BNN}_{\mathrm{2D}}^{\mathrm{Mem}_{-m}}$ model to a sample of 16 nearby SDSS clusters spanning a range of masses and dynamical states.
For this limited sample, the comparison with the adopted literature $M_{200}$ estimates provides no strong evidence for a non-zero mean logarithmic mass offset.
The observational comparison remains subject to differences between the idealised mock catalogues and real spectroscopic samples, including uncertainties in cluster centring, incomplete and non-uniform spectroscopic sampling, and survey selection effects.
These results motivate further applications of the framework to larger and more representative observational cluster samples.

\begin{acknowledgements}
      L.P.F. acknowledges support from the National Natural Science Foundation of China (NSFC; grant No. 12541302), the Innovation Program of Shanghai Municipal Education Commission (grant No. 2025GDZKZD04), and the China Manned Space Project (grant No. CMS-CSST-2025-A05). W.D. acknowledges support from the China Manned Space Project (grant Nos. CMS-CSST-2025-A02 and CMS-CSST-2025-A03). Z.J.L. acknowledges support from the National Natural Science Foundation of China (grant No. 12573009) and the scientific research grant from the China Manned Space Project (grant No. CMS-CSST-2025-A07).
\end{acknowledgements}

\bibliographystyle{aa}
\bibliography{reference}

\end{document}